\documentclass[final,1p,times]{elsarticle}
\usepackage{amssymb}
\usepackage{graphics}
\usepackage{graphicx}
\usepackage{epsfig}

\usepackage{amsthm}
\usepackage{amsmath}
\journal{Advances in Colloids and Interface Science}

\begin{document}

\begin{frontmatter}

%\linenumbers


%% \author{Name\cortext[cor1]{}\fnref{label2}}

\title{Nanoemulsion stability: experimental evaluation of the flocculation rate from turbidity measurements}


\author[ivic]{Kareem Rahn-Chique\corref{cor1}\fnref{label2}}
\ead{krahn@ivic.gob.ve}
\author[ual]{Antonio M. Puertas}
\author[ual]{Manuel S. Romero-Cano}
\author[ivic]{Clara Rojas}
\author[ivic]{German Urbina-Villalba}

\cortext[cor1]{Corresponding author}
\fntext[label2]{\footnotesize{Tel: +58-(212) 5041980. Fax: +58-(212) 5041915.}}
\address[ivic]{Instituto Venezolano de Investigaciones Cient\'ificas (IVIC). Centro de Estudios Interdisciplinarios de la F\'isica (CEIF), Carretera Panamericana Km. 11, Aptdo. 20632, Caracas, Venezuela.}
\address[ual]{Universidad de Almer\'ia (UAL), Departamento de F\'isica Aplicada, 04120-Almer\'ia, Spain.}

\begin{abstract}

The coalescence of liquid drops induces a higher level of complexity compared to the classical studies about the aggregation of solid spheres. Yet, it is commonly believed that most findings on solid dispersions are directly applicable to liquid mixtures. Here, the state of the art in the evaluation of the flocculation rate of these two systems is reviewed. Special emphasis is made on the differences between suspensions and emulsions. In the case of suspensions, the stability ratio is commonly evaluated from the initial slope of the absorbance as a function of time under diffusive and reactive conditions. Puertas and de las Nieves (1997) developed a theoretical approach that allows the determination of the flocculation rate from the variation of the turbidity of a sample as a function of time. Here, suitable modifications of the experimental procedure and the referred theoretical approach are implemented in order to calculate the values of the stability ratio and the flocculation rate corresponding to a dodecane-in-water nanoemulsion stabilized with sodium dodecyl sulfate. Four analytical expressions of the turbidity are tested, basically differing in the optical cross section of the aggregates formed. The first two models consider the processes of: a) aggregation (as described by Smoluchowski) and b) the instantaneous coalescence upon flocculation. The other two models account for the simultaneous occurrence of flocculation and coalescence. The latter reproduce the temporal variation of the turbidity in all cases studied (380 $\leq [NaCl] \leq $ 600 mM), providing a method of appraisal of the flocculation rate in nanoemulsions.

%The complexity of emulsions is significantly greater than the one of suspensions. Yet, it is commonly believed that some findings of solid dispersions are directly applicable to liquid mixtures. In this review the state of the art in the evaluation of the flocculation rate is presented. Special emphasis is made on the differences between suspensions and emulsions, and on the role of the surfactant molecules in the coupling of the mechanisms of Ostwald ripening, flocculation, coalescence and creaming. In the case of suspensions, the stability ratio is commonly evaluated from the initial slope of the absorbance as a function of time under diffusive and reactive conditions. Puertas and de las Nieves (1997) developed a theoretical approach that allows the determination of the flocculation rate from the variation of the turbidity of a sample as a function of time. Here, suitable modifications of the experimental procedure and the referred theoretical approach are implemented in order to calculate the values of the stability ratio and the flocculation rate corresponding to a dodecane-in-water nanoemulsion stabilized with sodium dodecyl sulfate. The agreement between theory and experiment is good, providing a method of appraisal of the flocculation rate of nanoemulsions in the presence of a significant amount of salt.

\end{abstract}


\end{frontmatter}

%\linenumbers

\begin{small}
\tableofcontents
\end{small}

% \linenumbers

\section{Introduction}

The absolute coagulation rate is an important parameter used to characterize the coagulation kinetics of colloidal systems. Knowledge of the stability and flocculation kinetics of these systems is essential for different applications in chemical, food, pharmaceutical, and material industries, environmental engineering, biotechnology, and nanotechnology \cite{Allain1995,Finsy1985,Stanislav2005,Sun2005}.

Smoluchowski \cite{Smoluchowski1917} considered the kinetics of irreversible flocculation of a homogeneous dispersion, initially composed of single particles suspended in a liquid. The number concentration of aggregates of $k$ primary particles (size $k$) existing in a dispersion at a given time ($n_{k} (t)$) results from a balance between the aggregates produced by the collisions of smaller clusters of sizes $i$ and $j$ ($i+j=k$), and the number of aggregates of size $k$ lost by a collision with aggregates of any other size:
\begin{equation}\label{Smoluchowski2}
    \frac{dn_{k}(t)}{dt} = \frac{1}{2} \sum_{i=1, \; j=k-i}^{i=k-1} k_{ij}\; n_{i}(t)n_{j}(t)-
    n_{k}(t) \sum_{i=1}^{\infty} k_{ik} \; n_{i}(t).
\end{equation}
The kernel of the Eq. \ref{Smoluchowski2} is the set of coagulation rates of aggregates of size $i$ and $j$, $\{k_{ij}\}$. Assuming equal flocculation rates for all aggregates of different size ($k_{ij}\,=\,k_{f}$), Smoluchowski deduced a simple analytical equation for the change in the total number of aggregates as a function of time ($n_{a}$):
\begin{equation}\label{Smoluchowski}
    n_{a}=\frac{n_{0}}{(1+k_{f}n_{0}t)}.
\end{equation}
Here, $n_{0}$ stands for the initial number of aggregates $(n_{a} (t = 0) = n_{0})$. The flocculation rate suggested by Smoluchowski, $k_{f}=4k_{B}T /3\eta$ (where: $k_{B}$, $T$ and $\eta$ stand for the Boltzmann constant, the absolute temperature and the viscosity of the external phase, respectively) solely considers the Brownian movement of the particles.

Derjaguin-Landau-Verwey-Overbeek \cite{Verwey1948,Verwey1946,Derjaguin1941}, introduced the effect of the attractive ($V_{A}$) and the electrostatic ($V_{E}$) potentials in the process of flocculation. According to their theory (DLVO), the total potential of interaction between the particles, $(V_{T}=V_{E}+V_{A})$ shows a primary minimum, a repulsive barrier, and a secondary minimum. Flocculation in the primary minimum is assumed to be irreversible due to the strength of the van der Waals interaction
at short distances. The onset of destabilization occurs when the maximum of the repulsive barrier of the potential is equal to zero ($V_{T}=0$ and $\frac{\partial V_{T}}{\partial r_{ij}}=0$, where $r_{ij}$ is the distance between the particles $i$ and $j$). Nonzero repulsive barriers provide kinetic stability to the system with respect to irreversible flocculation \cite{Chandrasekar1943,Kramer1943}.

To account for the effect of a repulsive barrier in the equation of Smoluchowski, Fuchs \cite{Fuchs1934} introduced the concept of a stability ratio, $W$. Later on, McGown and Parfitt \cite{McGown1967} modified it to include the attractive interaction in the calculation of fast aggregation, and Spielman \cite{Spielman1970} introduced a correction due to the hydrodynamic interaction:
\begin{equation}\label{W-Fuchs}
    W_{ij}= \frac{\displaystyle \int_{{R_{i}}+R_{j}}^{\infty} \; [f(r_{ij})/r_{ij}^{2}] \; exp(V_{T}/k_{B}T) \; dr_{ij}} {\displaystyle \int_{{R_{i}}+R_{j}}^{\infty} \; [f(r_{ij})/r_{ij}^{2}] \; exp(V_{A}/k_{B}T) \; dr_{ij}},
\end{equation}
where, $R_{k}$ is the radius of particle $k$. The pre-exponential factors of Eq. \ref{W-Fuchs} account for the effect of the hydrodynamic interactions on the particles movement.

\subsection{Flocculation of solid particles}

A comprehensive compendium of the early experimental techniques and theoretical approaches used to quantify the flocculation rate of suspensions was published by Sonntag and Strenge \cite{Sonntag1987} in 1987. A formal treatment of the hydrodynamic aspects of particle diffusion and electrokinetic effects can be found in the books of Russel et al. \cite{Russel1989}, van de Ven \cite{vandeVen1989}, Lyklema \cite{Lyklema1995}, Hunter \cite{Hunter2001}, Adamson \cite{Adamson1990}, Elimelech et al. \cite{Elimelech1995} and Dhont \cite{Dhont1996}. Up to the 1980's most of the research on this subject dealt with: a) the testing of constant-charge and constant-potential solutions of the Poisson-Boltzman equation required for the calculation of the electrostatic interaction between the particles, b) the appraisal of hydrodynamic interactions, c) the differences between the electrostatic potential of the particles at the interface, and the zeta potential deduced from mobility measurements, d) the inclusion of non-DLVO forces between the particles, e) the effect of counterions on the stability of the suspension, and f) the consequences of reversible (secondary minimum) flocculation \cite{Mamur1979,Behrens2000,Urbina2005}.

A review of the experimental methods currently employed for measuring aggregation rates was recently published by Gregory \cite{Gregory2009}. Lips et al. \cite{Lips1971}, and also Lichtenbelt et al. \cite{Lichtenbelt1973} developed experimental methodologies suitable for quantifying the rate of doublet formation ($k_{11}$) with light scattering. The procedure reported in Ref. [26] allows obtaining the value of $W$ from the initial slope of the turbidity as a function of time \cite{Lips1971,Lips1973,Lichtenbelt1973,Young1991}. This methodology is very convenient since it only requires the use of a spectrophotometer. Reering and Overbeek deduced an analytical relationship between $log\, W$ and $log\, I_{s}$ (where $I_{s}$ is the ionic strength of the solution) which was confirmed by the experiments \cite{Reering1954}. Using numerical evaluations of the interaction potentials, Prieve and Ruckenstein reported a linear dependence between $log\, W$ and the height of the repulsive barrier between the particles \cite{Prieve1980}.

In order to measure absolute coagulation rates, Young and Prieve \cite{Young1991} improved the small-angle scattering technique proposed by Lips and Willis \cite{Lips1973} shortening the path length of light through the dispersion. For rapid flocculation occurring at high salt concentration, a stability ratio of $2.14$ was obtained for a polystyrene latex of diameter $d=$486 nm.

Sun, Liu and Xu \cite{Sun2006,Liu2007,Xu2010} made an extensive study regarding the necessary conditions for a reliable turbidity determination of the coagulation rate. The relative rate of turbidity change at short times: $R_{T} = (1/\tau_{0})(d\tau/dt)_{0}$, is found to depend markedly on the wavelength of the incident light and the size parameter $\alpha = 2 \pi a / \lambda$ (where $a$ is a characteristic length of the particle). If the extinction cross section of two singlets ($2 \sigma_{1}$) is less than that of one doublet ($\sigma_{2}$), then $R_{T} > 0$, otherwise $R_{T} \leq 0$. Depending on $\lambda$, $R_{T}$ can be equal to zero (blind point) if $\sigma_{2} \sim 2 \sigma_{1}$. In this case, the turbidity is insensitive to the aggregation process. In mixtures of particles with average radii $R =$ 115, 266, and 500 nm, the blind point is found to depend on the size of the particles and on the fraction of each component \cite{Liu2007}. Furthermore, larger particles greatly mask smaller ones, inducing experimental errors.

These authors \cite{Sun2006,Liu2007,Xu2010} demonstrated that the T-matrix method can calculate the optical cross sections of singlets and doublets accurately, and confirmed that the Rayleigh-Gans-Debye (RGD) \cite{Kerker1969} approximation is only reliable for very small particles: $\alpha < 3$. When this condition is not fulfilled, the value of $k_{11}$ changes as a function of the wavelength. The value of $\lambda$ that maximizes $R_{T}$ guarantees better measurements, but it is restricted by the additional constraint on $\alpha$. Further uncertainties can be caused by unwanted contributions of the turbidity due to the forward scattering of light \cite{Xu2010}.

Toward the end of the last century, the synthesis of functionalized latex particles of micron-size became widespread as well as the use of static (SLS) and dynamic (DLS) light scattering techniques. The latex particles have the size of a few hundred nanometers, and are almost neutrally buoyant. As a result of these advances, the morphology of the aggregates was linked to the aggregation rate \cite{Lin1989,Meakin1992,Feder1984}, and some mathematical properties of the flocculation rate were quantified thanks to the dynamic scaling of the population balance equation for irreversible aggregation \cite{Feder1984,Ball1987,Lin1990,Vicsek1984,Manley2004,Weitz1984,Sandkuhler2005}. From these studies, a general form of the flocculation rate between aggregates of sizes $i$ and $j$ was deduced \cite{Lattuada2003}:
\begin{equation}\label{kij}
    k_{ij}=\left(\frac{K}{W_{ij}}\right) B_{ij} P_{ij}.
\end{equation}
Here $K$ is the kernel for Brownian aggregation, $B_{ij}$ is a geometric correction which accounts for the fractal dimension of the aggregates, $W_{ij}$ is the stability ratio between particles $i$ and $j$, and $P_{ij}$ is the variable that contains the details of the aggregation mechanism.

Using single particle light scattering (SPLS), Holthoff et al. \cite{Holthoff1996} studied the time evolution of aggregates from doublets up to heptamers. For particles of $d=$580 and 683 nm, and $I_{s} = 1$ M KCl, the experimental data nicely fitted the predictions of Smoluchowski corresponding to a constant kernel, with $k_{11}$ = $(7.3-7.1)$ $\times 10^{-18} m^{3} s^{-1}$ \cite{Holthoff1997}.  Similar values were obtained using DLS and SLS techniques $(6.9-6.1)$ $\times 10^{-18}$ $m^{3} s^{-1}$.  Moreover, a linear variation of $log\, W$ vs. $log \,I_{s}$ was found \cite{Holthoff1996}. For a salt concentration of 250 mM, the stability ratio varied in the range of 2.5-2.8 ($11.9 < W < 14.2$ for $I_{s} =$ 125 mM). Using DLS and SLS, Behrens et al. \cite{Behrens1998} measured the absolute coagulation rate of doublet formation and compared it with the theoretical predictions of the standard DLVO theory. The dependence of the stability ratio with the pH, the surface charge of three particles (two carboxylic lattices ($d=$309, 104 nm) and hematite ($d=$122 nm)), and the ionic strength of the solution was studied. For the systems under consideration the predictions of the formulas corresponding to constant charge, constant potential and regulated charge do not differ much. It was concluded that for colloid particles of low surface charge ($< 3\, mC \, m^{-2}$) and moderately low ionic strength ($0.1-10$ mM) the classical DLVO theory works quantitatively. However, when the ionic strength is increased to 100 mM, the maximum of the potential curve lies below 2.2 nm, and a deep secondary minimum develops. In this case, the slope of $W$ vs. $pH$ predicted by the calculations differs considerably from the one suggested by the experimental measurements. For these experiments, the value of $W$ changed between 1 and 1000, and the fast flocculation rate for doublet formation ($k_{11}^{fast}$) ranged from 2.4 $\times 10^{-18}$ to 3.5 $\times 10^{-18} m^{3} s^{-1}$.

Nowadays, SLS and DLS techniques are commonly applied to study homo and hetero-aggregation in complex systems \cite{Yu2002,Puertas1999,Schmitt2001}. These techniques allow obtaining the value of $k_{11}$ under reactive and diffusive conditions for solid particle dispersions. However, bigger aggregates are almost exclusively considered for topological studies, not for the calculation of the aggregation rate.

\subsection{Flocculation of oil drops}

Typical macro and nanoemulsions are kinetically stabilized dispersions of oil (o) in water (w). Unlike suspensions they are subject to coalescence, Ostwald ripening and phase inversion, besides flocculation and creaming. These phenomena occur simultaneously and are not independent of each other. Emulsion drops are usually of micrometer size and their coalescence dynamics includes the occurrence of deformation, capillary waves, and hole nucleation. The density of the oils varies between 0.6 and 0.9 $g \, cm^{-3}$, which generates a substantial amount of creaming upon aggregation and particle growth. For an overview of some of the problems related with emulsion stability, the reader is referred to the following books: \cite{Petsev2002libro,Sjoblom2001libro,Binks1998libro,Ivanov1998libro}.

While big drops ($R > 50 \mu m$) deform during flocculation, small drops ($R < 1 \mu m$) are reluctant to deform due to their high Laplace pressure \cite{Rojas2010}. Whenever deformation occurs the draining of the intervening film between the drops takes a finite time \cite{Ivanov1999,Danov1993,Couder2005}. Although the process of coalescence can be formidably complex \cite{Velikov1997}, deformation allows distinguishing between flocculation and coalescence. However, if two drops do not deform during their approach, the final state of the system depends on the interaction potential. A high repulsive barrier will promote secondary minimum flocculation as in the case of solid particles, but a small or non-existent barrier, will lead to a steady approach until coalescence occurs. In this case, it is not possible to determine when the process of flocculation finishes and the one of coalescence begins. Hence, an experimental quantification of a flocculation rate under diffusive conditions necessarily includes the process of coalescence \cite{Osorio2011}.

The deformation of the drops depends on their size, hydrodynamic interactions, colloidal forces, and surface forces \cite{Ivanov1999,Tabor2011}. According to recent emulsion stability simulations (ESS) an approximate threshold for drop deformation occurs around 2.5 $\mu m$ \cite{Rojas2010,Ivanov1999,Basheva1999,Dickinson1988}. Drops with radii smaller than 2.5 $\mu m$ will mostly behave as non-deformable particles.

In the absence of surfactants, oil drops show a substantial electrostatic potential. This surface charge is commonly ascribed to the preferential adsorption of hydroxyl groups at the oil/water interface \cite{Stachurski1996,Marinova1996,Beattie2004}. Hence, it depends on the pH and on the total interfacial area of the emulsion.

Using DLS, Sakai et al. \cite{Sakai2003} observed that the radii of benzene droplets dispersed in water increase in a geometrical series with a common ratio of ($V/ 3V_{o})^{1/3}$, where $V$ is the total volume of the emulsion ($V=V_{w}+V_{o}$), and $V_{i}$ is the volume of phase $i$. This result was explained supposing a multi-step-growth model which assumes that primary droplets first aggregate into a cluster, and then merge (coalesce) making a secondary droplet. Similarly, secondary droplets aggregate and then merge into a tertiary droplet, and so on. The rate of the process depends on the mean free path of the drops.

Sakai et al. \cite{Sakai2001} studied the evolution of alkane-in-water nanoemulsions ($R=$ 30 nm, $\phi=$ 2 $\times 10^{-4}$, where $\phi$ is the volume fraction of oil) in the absence of surfactants. According to the theory of Lifshitz, Slyozov and Wagner (LSW theory) \cite{Lifshitz1961,Wagner1961}, the Ostwald ripening rate ($dR^{3}/dt$) is proportional to the solubility of the oil, and the interfacial tension of the drops. Our group carried out simulations on the behavior of the dodecane-in-water nanoemulsions studied by Sakai \cite{Urbina2009a}. In the absence of salt, the simulations reproduced the experimental behavior, showing that: a) the initial evolution of $R^{3}$ is only due to flocculation and coalescence, b) the slope of $R^{3}$ vs. $t$ decreases with time approaching the LSW limit, and c) molecular diffusion only contributes to the augment in $R$ after 800 s. At high salt concentration it was predicted that $R^{3}$ vs. $t$ varies linearly as a product of flocculation and coalescence producing a polydisperse emulsion.

Most of the works on emulsion stability concern macroemulsions. Using video enhanced microscopy, Holt et al. \cite{Holt1997,Holt1998,Dukhin2003} studied the lifetime of a doublet ($t_{d}$) in toluene in water emulsions ($\phi=$ 0.005-0.02, $R \sim$ 2-3 $\mu m$) at low ionic strength (3-13 mM NaOH).  At $I_{s} <$ 3 mM there is no influence of $I_{s}$ over $t_{d}$. At $I_{s} >$ 3 mM, the distribution of lifetimes splits into two parts: one at small contact times that do not depend on the value of $I_{s}$, and another at large lifetimes which grow longer as the electrolyte concentration increases. It was found that the former lifetimes correspond to collisions between two particles in which the minimum distance of separation is longer that the position of the secondary minimum. The latter values of $t_{d}$ correspond to secondary minimum flocculation. The depth of the minimum augments as the ionic strength rises, increasing the value of $t_{d}$. Values of $t_{d}$ of the orders of several minutes were observed, as well as evidence of droplet coalescence inside doublets. For very high ionic strength ($I_{s} =$ 0.1 M NaOH, $W \sim2$), the decrease in the number of singlets closely follows the predictions of Smoluchowski. It was also observed that the drops tend to stick to the walls of the coalescence channel when $I_{s}\geq$ 0.1 M.

Verbich et al. \cite{Verbich1997} studied the evolution of a 1,1-dichlorodecane in water macroemulsion ($\phi=$ 0.01, $R\sim$ 2-4 $\mu m$, $10^{-4}\leq$ [NaOH] $< 10^{-2}$ M). They found that the relative absorbance, $Abs/Abs_{0}$ decreased with time during several days. The zeta potential of the drops also decreased unless SDS was added to the system. The value of the stability ratio was calculated using the time required for a given emulsion ([NaOH] $= 10^{-3}$ M) to reach the same optical density of a fast flocculating one ([NaOH] $= $0.3 M) $W \sim 48$.  The emulsions did not show a sensitive variation of coagulation rate with the ionic strength, which does not agree with the theory of Fuchs. Using optical microscopy it was observed that as the drops of 2 $\mu m$ disappear, the content of 6-10 $\mu m$ drops increased gradually, and larger drops start to appear. Finally, the number of doublets consisting in one large droplet (12-16 $\mu m$) and one small droplet increased. Multiplets were observed after several days.

Karaman et al. \cite{Karaman1996}, Pashley \cite{Pashley2003}, and Maeda et al. \cite{Maeda2004} suggested that dissolved gas has a key role in the stability of surfactant-free dodecane-in-water emulsions. Degassed oil spontaneously emulsifies in water producing stable emulsions whose turbidity progressively decreases as a function of time. In apparent contradiction, the reintroduction of dissolved gases does not destabilize the emulsions already formed. The turbidity of emulsions not degassed was considerably lower, but also diminished with time. This was ascribed to the buoyancy removal of the large drops. Burnett et al. showed that the behavior of these emulsions is related to the freeze-pump-thaw treatment and not to the amount of gas dissolved \cite{Burnett2004,Eastoe2007}.

Ionic surfactants adsorb to the interface of the drops increasing their electrostatic charge \cite{Marinova1996,Barchini1996}. The charge depends on the equilibrium adsorption of the surfactant, which is a function of the surfactant concentration. At a certain concentration surfactants aggregate into micelles. This equilibrium is modified by the presence of salt in the aqueous media that decreases the critical micelle concentration and increases the surfactant surface excess. However, as the ionic strength increases, the additional charge is also partially screened by the salt counterions.

High salt concentrations decrease the solubility of the surfactant in the aqueous phase. The surfactant may precipitate as hydrated crystals if $\phi$ is low \cite{Wennerstrom}. In the most general case, high salt concentrations favor the formation of a separate surfactant-rich middle phase, or even the transfer of the surfactant to the oil phase in the form of inverse swollen micelles \cite{Aveyard1985,Salager1979,Shinoda1967,Salager1982}. The region of compositions for which a three-phase system exists at equilibrium is known as the balance zone. Within this region the interfacial tension produced by gemini surfactant attains ultralow values. Aveyard et al. \cite{Aveyard1987} showed that a single chain surfactant like SDS also generates a tension minimum in the balance zone, but it cannot reach ultralow values of the tension unless a significant amount of a cosurfactant is added to the system. Interestingly, the stability of emulsions in the balance zone is also minimum, changing in several orders of magnitude within a narrow region of compositions \cite{Osorio2011,Kabalnov1996,Kabalnov1996a,Binks2000}. This lack of stability has been the subject of several hypotheses. Recent works connect unstable emulsions with the spontaneous curvature of the surfactant at the interface, and the pronounced occurrence of Ostwald ripening. If the salt concentration is increased beyond the balance zone, the excess surfactant is transferred to the oil phase and the emulsion passes from $o/w$ to $w/o$ \cite{Aveyard1985,Salager1979,Shinoda1967,Wennerstrom}.

Using silicon-in-water emulsions ($0.34 < R < 3.3$ $\mu m$) stabilized with sodium dodecyl sulfate (SDS), Bibette et al. \cite{Bibette1992} demonstrated that a sharp destabilization threshold can be defined. In these experiments, dialysis bags and Dextran were used to set the value of the osmotic pressure. The threshold was found to be a function of the surfactant concentration, the osmotic pressure, and the droplet diameter.

Schmitt et al. \cite{Schmitt2004,Schmitt2004a} studied the behavior of concentrated macroemulsions of alkane-in-water, noticing that Ostwald ripening dominates the evolution of the emulsion shortly after emulsification. However, a few hours later, coalescence increases abruptly equalizing and surpassing the Ostwald ripening rate. The point at which the two rates of destabilization are equal, allows the quantification of the frequency of rupture of $o/w/o$ films, which is a measure of the coalescence frequency. This frequency is around $10^{4}$ $m^{-2}s^{-1}$ for the case of heptane-in-water emulsions stabilized with SDS.

Aveyard et al. \cite{Aveyard2002,Urbina2006} and Binks et al. \cite{Binks2000} studied the flocculation transitions of dodecane-in-water and heptane-in-water macroemulsions ($3 < R < 10$ $\mu m$) stabilized with non-ionic surfactants. These emulsions are stable with respect to coalescence and Ostwald ripening, but unstable with respect to flocculation and creaming. The transitions between non-flocculated and flocculated states were induced by the addition of an ionic surfactant and salt. In this case, the surface charge of the drops depends on the partition of the ionic surfactant between the aqueous solution and the interface. Using analytical forms for each potential contribution, the total interaction energy and the disjoining pressure between two deformable emulsion drops were calculated. This allowed the definition of regions of flocculation in plots of the ionic surfactant concentration vs. the total ionic strength. Additionally, the destabilization of the emulsion was quantified in terms of the creaming rate. The comparison between theory and experiment was very good.

van den Tempel considered the process of coalescence of $o/w$ systems as a first order reaction depending on the number of films between aggregated drops \cite{vandenTempel1953}. Using the equation of Smoluchowski and assuming linear aggregates, he was able to deduce an explicit analytical expression for the total number of aggregates in an emulsion, which depends on a flocculation and a coalescence rate. Borwankar et al. deduced a similar average expression for the number of aggregates, considering the coalescence of small flocs \cite{Borwankar1992}. Fitting the theoretical expressions to the experimental data from soybean in water emulsions ($\phi=$ 0.30), flocculation rates of the order of $10^{-16}$ $m^{3} s^{-1}$ were obtained, along with coalescence rates between 7.4 $\times 10^{-6}$ and 2.5 $\times 10^{-4} s^{-1}$.

Reddy, Fogler and Melik \cite{Reddy1981,Melik1984,Melik1985} studied the simultaneous occurrence of flocculation and creaming by numerical evaluation of the respective population balance equation. This procedure allowed explaining the cyclic shifts of the drop size distribution (DSD) toward higher and lower particles sizes, observed in paraffin/water emulsions ($R\sim1-4$ $\mu m$). Moreover, they reproduce the evolution of the DSD at different heights of the container. These distributions differ significantly from one another.

A kinetic model for the simultaneous processes of flocculation and coalescence was forwarded by Danov et al. \cite{Danov1994}. Most of their differential equations must be evaluated numerically. However, they demonstrated that if the coalescence rate of an emulsion is significantly faster than the flocculation rate, and big drops are counted as single particles, the expression of Smoluchowski for the total number of aggregates is recovered. This result was repeatedly validated by ESS of non-deformable drops \cite{Urbina2004,Urbina2005a}. In fact, an alternative expression for the total number of aggregates was deduced \cite{Urbina2004}. This expression is valid for systems of distinct polydispersity with 0.01 $<\phi<$ 0.30, $10^{-5}$ M $< C_{surf} <$ 8 $\times 10^{-5}$ M (where $C_{surf}$ is the surfactant concentration). In the absence of a repulsive barrier, limiting values for the mixed flocculation-coalescence rate ($k_{FC}$) of non-deformable drops were also calculated \cite{Urbina2004}. The value of $k_{FC}$ depends on the volume fraction of oil and the Hamaker constant ($A_{H}$) of the van der Waals potential between the drops. At infinite dilution $k_{FC}$ is equal to 6.4 $\times 10^{-18}$ $m^{3} s^{-1}$ for $A_{H} =$ 1.24 $\times 10^{-21}$ J, and 9.5 $\times 10^{-18}$ $m^{3} s^{-1}$ for $A_{H} =$ 1.2 $\times 10^{-19}$ J.

The aggregation dynamics of non-deformable \cite{Urbina2000,Urbina2005,Urbina2005a,Urbina2006,Urbina2009,Urbina2004,Urbina2009a} and deformable \cite{Rojas2010,Rojas2010a,Osorio2011,Toro-Mendoza2010,Urbina2009} drops subject to coalescence has been extensively studied by our group using simulations. Since the scale of drops and surfactants differs in several orders of magnitude, the explicit movement of both entities at the same time is not yet computationally feasible.  Hence, the major problem of ESS is to consider surfactant effects without moving the surfactant molecules explicitly. The second major difficulty is to account for the influence of a finite surfactant population in all destabilization processes simultaneously.  The surfactant surface excess depends on time due to the continuous decrease of the available interfacial area caused by the coalescence of the drops, and the molecular diffusion of the amphiphile \cite{Urbina2004a}. The adsorption of surfactant  determines the interfacial properties of the drops such as tension, charge, fluid mobility, etc. These properties change the deformability of the drops, as well as their interaction potentials, hydrodynamic tensors, and the coalescence mechanism (film drainage, capillary waves, hole formation) \cite{Rojas2010}. Hence, all destabilization processes depend on the surfactant distribution throughout the system, and this distribution is continuously changing.

Former ESS calculations were exclusively devoted to non-deformable drops and ionic surfactants. Although ESS confirmed most of the predictions of the DLVO theory, some important differences appeared. According to the simulations of nondeformable drops, the stability threshold occurs as soon as the maximum of the repulsive potential starts to form \cite{Urbina2009}: $\frac{\partial V_{T}}{\partial r_{ij}}=0$ and $\frac{\partial^{2} V_{T}}{\partial r_{ij}^{2}}=0$. The behavior of the systems depends strongly on the surfactant concentration and the total interfacial area of the emulsion. If the surface excess of the surfactant is approximately constant, as it occurs for very low or very high surfactant concentrations, DLCA (Diffusion Limited Cluster Aggregation) and RLCA (Reaction Limited Cluster Aggregation) regimes are observed, respectively \cite{Urbina2006}. However, if the redistribution of surfactant molecules after a coalescence event is considered, the surface potential of the drops evolves dynamically, promoting a progressive stabilization of the emulsion with respect to flocculation and coalescence \cite{Urbina2009}. In these cases, the fractal coefficient of the aggregates changes with time, and the equation of Smoluchowski has to be modified to account for the total number of aggregates. Some recent experimental evidence on the existence of surfactant redistribution upon coalescence can be found in Refs. \cite{Ghosh2002,Hu2003,Grimes2010,Gauer2009,Gauber2010}.

Using silicon-in-water nanoemulsions of 276, 374 and 576 nm and SDS concentrations of 3, 5, and 7 mM, Barchini and Saville \cite{Barchini1996} showed that the zeta potential predicted by mobility and conductivity measurements cannot be reconciled. The system of 576 nm with 5 mM even shows a negative slope of the conductivity as a function of the volume fraction of oil ($\phi < 0.05$). These inconsistencies were ascribed to the inadequacy of the standard electrokinetic theory to describe the high surface potentials of the droplets and the properties of the inner part of the diffuse double layer. However, recent evidence indicates that the interfacial area occupied by the surfactant in nanoemulsions could be one order of magnitude higher than the one expected from the equilibrium adsorption of the surfactant to macroscopic interfaces \cite{Aguiar2011}.

Poulin et al. \cite{Poulin1999} noticed that the behavior of silicon-in-water nanoemulsions ($R=$ 190 nm, [NaCl]$ =$ 0.35, [SDS] $=$ 2mM) depends on the temperature and on the volume fraction of oil. Following the changes of the scattering peak, the aggregation mechanism was identified as DLCA for deep quenches and a volume fraction of oil between $0.01 <  \phi < 0.1$. For $\phi > 0.1$ and shallow quenches, different mechanisms closer to spinodal decomposition were observed. It was also noticed that the droplets become highly adhesive at high salinity exhibiting large contact angles. According to Wang et al. \cite{Wang2009}, flocculation is the main mechanism of destabilization in water/decane emulsions stabilized with a mixture of pentaoxyethylene lauryl ether ($C_{12}E_{5}$) and di-dodecyldimethylammonium bromide.

As any $o/w$ emulsion, nanoemulsions are subject to creaming and Ostwald ripening \cite{Kabalnov1987,Weers1999,Taylor1995,Kabalnov1990,Weiss1999}. However, nano-metric particles should not be appreciably subject to creaming unless they aggregate. In the absence of salt, all systems investigated showed an appreciable amount of creaming after two days. Creaming might be prevented mixing dodecane with a high-density oil, to make the mixture neutrally buoyant. Unfortunately, most high density liquids are halogenated alkanes, whose solubility in water is higher than the one of dodecane. This favors the occurrence of Ostwald ripening. The theoretical ripening rate of surfactant-free dodecane/water emulsions is equal to 1.3 $\times 10^{-26} m^{3} s^{-1}$. However, this rate changes with surfactant addition. Ostwald ripening might be prevented mixing dodecane with an insoluble oil \cite{Kamogawa1999,Delmas2011}. A systematic study on the influence of creaming and ripening on the value of the flocculation rate is currently carried on in our lab. In this article we concentrate on the evaluation of aggregation in a typical nanoemulsion.

It is evident from above that emulsions are far more complex than suspensions. As a result, the quantification of the processes leading to destabilization is more difficult, even if the relevant variables involved are clearly identified. The prediction of the temporal evolution of these systems requires elaborate simulation techniques \cite{Urbina2009,Urbina2009a,Rojas2010,Rojas2010a}, and the quantification of each destabilization process is restricted to those conditions in which the rate of each mechanism is substantially different from the rest. Moreover, it is uncertain whether these rates are relevant in typical systems where all processes occur simultaneously, regulated by a time-dependent surfactant distribution.

In this work, the initial 30-seconds variation of the turbidity of an ionic dodecane-in-water nanoemulsion is used to evaluate its flocculation rate at different ionic strengths. This time span is considerably larger than the one required for the formation of doublets. The nanoemulsions studied were prepared by the phase inversion composition method \cite{Maestro2008, Forgiarini2001,Wang2008,Sole2006}, which basically consists in the sudden dilution of a pre-equilibrated system in the balance zone. The experimental determination of $W$ as a function of the initial slope of the turbidity ($\tau$) vs. time ($t$), is complemented with four theoretical models of the turbidity which depend on the average drop size and the flocculation rate. These models are approximate formalisms which allow quantifying the stability of nanoemulsions with respect to flocculation and coalescence. Adjustment of the theoretical expressions to the experimental data allows the evaluation of either a flocculation rate or a mixed flocculation-coalescence rate. This methodology provides a complementary method to evaluate $W$, which proves to be in agreement with direct experimental determinations.

In the following section, the necessary theoretical background of the work is presented. In the third section the experimental procedure is outlined. The fourth section contains the results of the measurements and the theoretical fittings. In the fifth section, the results are discussed. The paper finishes with some concluding remarks.

\section{Theoretical background}

According to the scheme proposed by Smoluchowski, the process of aggregation begins with the formation of doublets:
\begin{equation}\label{k11}
    \frac{dn_{1}}{dt}=-k_{11}n_{1}^{2}-k_{12}n_{1}n_{2}\,-...
\end{equation}
\begin{equation}\label{k11(2)}
    \frac{dn_{2}}{dt}=\frac{k_{11}}{2}n_{1}^{2}-k_{12}n_{1}n_{2}\,-...
\end{equation}
where $n_{1}$ and $n_{2}$ are the number concentration of single and double particles, and $k_{11}$ and $k_{12}$ are rate constants.

In general, the turbidity of a colloidal dispersion can be defined as \cite{Lips1971,Lips1973}:
\begin{equation}\label{Turbidity}
    \tau=\sum_{k=1}^{\infty} n_{k}(t)\sigma_{k},
\end{equation}
where $n_{k}$ is the number of aggregates of $k$ particles per unit of volume, and $\sigma_{k}$ is their total light scattering cross section. Differentiation of Eq. \ref{Turbidity} gives:
\begin{equation}\label{dtau}
    \frac{d\tau}{dt}=\sigma_{1} \frac{dn_{1}}{dt} + \sigma_{2} \frac{dn_{2}}{dt}+...
\end{equation}
For the limiting case of $t\longrightarrow 0$, $n_{2}$ approaches zero and the second and further terms on the right side of Eqs. \ref{k11}, \ref{k11(2)}, and \ref{Turbidity} can be disregarded. Combining Eqs. \ref{k11}, \ref{k11(2)}, \ref{Turbidity}, and \ref{dtau}:
\begin{equation}\label{Constante-dimeros}
    \bigg(\frac{d\tau}{dt}\bigg)_{0}=230\bigg(\frac{dAbs}{dt}\bigg)_{0}=\bigg(\frac{1}{2}\sigma_{2}-\sigma_{1}\bigg) k_{11} n_{0}^{2}.
\end{equation}
Here, $\sigma_{1}$ and $\sigma_{2}$ are the optical cross-sections of a single spherical particle and a doublet, $\tau= (1/L) \ln(I_{0}/I)$, $Abs=\log(I_{0}/I)$ is the absorbance of the dispersion, $I$ is the intensity of light emerging from a cell of width $L$ ($10^{-2}m$), and $I_{0}$ is the intensity of the incident light.

The above cross sections can be evaluated using a suitable theory of light scattering. In this case, $k_{11}$ can be deduced from turbidity measurements.

Rayleigh, Gans, and Debye \cite{Kerker1969} formulated an approximate theory of light scattering, restricted to particles with small relative refractive index, and small size compared to the wavelength of the incident light. This theory is valid whenever:
\begin{equation}\label{RGD}
    C_{RGD}=(4\pi a/\lambda)\, (m-1)\ll 1.
\end{equation}
Here, $\lambda$ is the wavelength of light in the medium, and $m$ is the relative refractive index between the particle and the medium.

Kerker \cite{Kerker1969} estimated the regions of $C_{RGD}$ vs. $\alpha$ for which the error in the extinction cross section of  a sphere is lower than $10\%$. For latex particles ($m=$1.20), $\alpha$ must be lower than 2.0. In the case of dodecane ($m=$1.07, $\lambda=$400nm, $a=R=$72.5 nm) $\alpha=$1.51, and $C_{RGD}=$0.22. Hence, the errors produced in the calculation of the optical cross section of the drops using RGD are expected to be smaller.

$W$ can be obtained from the initial slope of the curve $\tau$ vs. $t$ at low and high ionic strength (slow and fast aggregation, respectively) \cite{Urbina-Villalba2009,Romero-Cano2000,Romero-Cano2001,Peula1997,Lichtenbelt1973,Gregory2009}:
\begin{equation}\label{W-turbidez}
    W\,=\,W_{exp}\;=\;\frac{\left(d\tau/dt\right)_{0,fast}} {\left(d\tau/dt\right)_{0,slow}}=\frac{\left(dAbs/dt\right)_{0,fast}}{\left(dAbs/dt\right)_{0,slow}}.
\end{equation}
Because in the case of solid particles the optical factors of Eq. \ref{Constante-dimeros} are canceled yielding:
\begin{equation}\label{W-turbidez2}
    W\,=\,W_{k_{11}}\;=\;\frac{k_{11}^{fast}}{k_{11}^{slow}}.
\end{equation}

It is also possible to fit the experimental data on the total number of aggregates to Eq. \ref{Smoluchowski} to obtain the value $k_{f}$. In this case, the stability ratio can be written in terms of the values of the flocculation rate in the absence ($k_{f}^{fast}$) or presence ($k_{f}^{slow}$) of a repulsive force:
\begin{equation}\label{W-experimental}
    W= \frac{k_{f}^{fast}}{ k_{f}^{slow}}.
\end{equation}
Thus, if the flocculation rates of aggregates of different size are similar, the average effect of the repulsive barrier can be incorporated in Eq. \ref{Smoluchowski} substituting $k_{f}$ by $k_{f}^{slow}=k_{f}^{fast}/W$.

\subsection{Theoretical models for the estimation of the turbidity curve}

\subsubsection{Aggregation of emulsion drops (model M1)}

Puertas et al. \cite{Puertas1997,Puertas1998} developed a general method for evaluating the average cross-section of an aggregate of size $k$. The total cross-section of each aggregate results from four contributions which take into account the possible structures that might be generated when each pair of particles in an aggregate is separated by zero, one or two particles between them.  Within the RGD approximation, the cross-section of an aggregate of $k$ spheres of radius $a$, can be expressed as \cite{Benoit1962,Kerker1969}:
\begin{equation}\label{sigmak}
    \sigma_{k}=\frac{4}{9}\pi a^{2} \alpha^{4} (m-1)^{2} \int_{0}^{\pi} P_{k}(\vartheta) (1+\cos^{2} \vartheta )\sin (\vartheta) d\vartheta.
\end{equation}
Here $\vartheta$ is the scattering angle, and
$P_{k}(\vartheta)$ is defined as:
\begin{equation}\label{Pk}
    P_{k}(\vartheta)=P_{1}(\vartheta)(k+A_{k}),
\end{equation}
where $P_{1}$ is given by:
\begin{equation}\label{P1}
    P_{1}(\vartheta)=\bigg(3\frac{\sin u-u \cos u}{u^{3}}\bigg)^{2},
\end{equation}
with: $u=2\alpha \sin (\vartheta/2)$, and
\begin{equation}\label{Ak}
    A_{k}=\sum_{i,j; \; i\neq j}^{k} \frac{\sin (hs_{ij})}{hs_{ij}}.
\end{equation}
Here, $s_{ij}$ is the center-to-center distance between spheres $i$ and $j$, and $h = u/a = (4\pi/\lambda) sin(\vartheta/2)$.

\begin{figure}[htbp]
\centering
  \includegraphics[width=9cm]{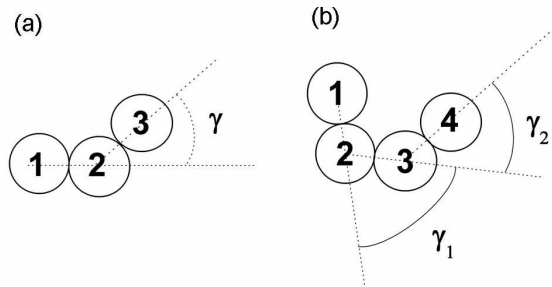}\\
  \caption{Definition of $\gamma$ for a chain of three particles (a) and four particles (b).}\label{Agregados2}
\end{figure}

For the scattering cross-section of aggregates with $k>4$, Puertas et al. considered in the above summation only the terms corresponding to pairs of particles separated by two other particles at most, because they decrease with the distance between them (note that pairs of particles close to each other but not connected by two particles or less are thus not accounted, which restricts the validity of the calculation to small aggregates). Following this procedure, the average value of $A_{k}$ can be calculated if it is considered that all structures of $k$ particles are equally probable. In this case:
\begin{eqnarray}\label{promedioAk}
    <A_{k}>=\sum_{i,j; \; i\neq j}^{k} \bigg < \frac{\sin(hs_{ij})}{hs_{ij}}\bigg > \approx \nonumber\\
   2(k-1)\frac{\sin(2ha)}{2ha}+2(k-2) \frac{3}{2\pi} \int_{0}^{2\pi/3} \frac{\sin(4u \cos (\gamma/2))}{2u\cos (\gamma/2)} d\gamma \nonumber\\
   + 2(k-3) \frac{9}{16\pi^2} \int_{-2\pi/3}^{2\pi/3} d\gamma_{1} \int_{-2\pi/3}^{2\pi/3} d\gamma_{2} \frac{\sin(hl(\gamma_{1}, \gamma_{2}))}{hl(\gamma_{1}, \gamma_{2})},
\end{eqnarray}
with $l(\gamma_{1},\gamma_{2})$ equal to:
\begin{equation}\label{funcionl}
l(\gamma_{1},\gamma_{2})=2a\left[1+2\cos (\gamma_{1}) + 2\cos (\gamma_{2}) + 4\cos^{2}\frac{(\gamma_{1}+\gamma_{2})}{2}\right].
\end{equation}
Angles $\gamma$, $\gamma_{1}$ and $\gamma_{2}$ are shown in Fig. \ref{Agregados2}.

Eqs. \ref{promedioAk} and \ref{funcionl} can also be calculated using an alternative set of equations which is formally equivalent, but more computationally efficient. These equations are given in Appendix A for further reference. Using the expressions of Smoluchowski for the kinetics of aggregation, the turbidity can be written as:
\begin{equation}\label{TurbidezPuertas}
      \tau_{M1}=\sum_{k=1}^{\infty} n_{k}(k_{f},t)\sigma_{k,a},
\end{equation}

where $\sigma_{k,a} = \sigma_{k}$ is the optical cross section of the aggregates, and:

\begin{equation}\label{nkPuertas}
      n_{k}=\frac{n_{0}(k_{f}n_{0}t)^{k-1}}{(1+k_{f}n_{0}t)^{k+1}}.
\end{equation}

Puertas and de las Nieves \cite{Puertas1997} used the kinetic constant, and the time $t_{0}$ as fitting parameters. Notice that $t_{0}$ is the time at which the process of aggregation begins. Theoretical calculations were compared with experimental evaluations of the aggregation of latex suspensions. A good agreement between theory and experiment was found. The theoretical approach reproduced the experimental data of $\tau$ corresponding to $50$ seconds of aggregation. The kinetic constants obtained agreed with the ones calculated from the initial slope of $\tau$ vs. $t$, and the literature ($k_{11}=2k_{f}$), indicating that the kinetics of Smoluchowski is a good approximation to the real aggregation process.

Thus, the model M1 refers to the direct application of this methodology to the case of emulsions. Despite the fact that the model developed in Refs. \cite{Puertas1997,Puertas1998} describes the aggregation behavior of a monodisperse system, we aim to apply it to a more real system, where polydispersity is a natural feature. It is clear that any distribution of particle size has a certain degree of polydispersity. However, the theory of Smoluchowski addresses the case in which all particles have the same radius at the beginning of the process. The fact that this formalism is commonly validated using suspensions (despite the inherent polydispersity of the systems) is an implicit recognition that the optical techniques accept a certain degree of tolerance in the particle size.

\subsubsection{Complete coalescence of the drops after contact (model M2)}

Model M2 assumes that the drops do not aggregate, but instead coalesce as soon as they collide with one another.

Drops coalesce if either the attractive potential or the thermal interaction with the solvent provides enough energy to overcome their repulsive interaction. In this case, a new particle with radius $a_{k}$ is created at the center of mass of the coalescing drops. Using the conservation of volume, $a_{i}^{3}+a_{j}^{3}=a_{k}^{3}$, hence,
\begin{equation}\label{ak}
    a_{k}=\sqrt[3]{k} a.
\end{equation}

Since, the form factor of a single sphere ($P_{k,s}$) of a radius $a_{k}$ is defined as:
\begin{equation}\label{PS}
    P_{k,s}(\vartheta)=\left (3\frac{\sin u_{k}-u_{k} \cos u_{k}}{u_{k}^{3}}\right)^{2}.
\end{equation}

The cross-section of a sphere of radius $a_{k}$ ($\sigma_{k,s}$) is given by:
\begin{equation}\label{sigmask}
    \sigma_{k,s}=\frac{4}{9}\pi a_{k}^{2} \alpha_{k}^{4} (m-1)^{2} \int_{0}^{\pi} P_{k,s}(\vartheta) (1+\cos^{2} \vartheta) \sin \vartheta d\vartheta,
\end{equation}
where $\alpha_{k}=2\pi a_{k}/\lambda$, and $u_{k}=2\alpha_{k}\sin (\vartheta/2)$.

If the drops do not deform, but coalesce as soon as they flocculate in the primary minimum, the turbidity can be calculated as:
\begin{equation}\label{TurbidityM2}
    \tau_{M2}=\sum_{k=1}^{\infty} n_{k}(k_{f},t)\sigma_{k,s}.
\end{equation}

This model is meant to reproduce the cases in which there is neither a barrier for flocculation or coalescence. Previous emulsion stability simulation demonstrated that in this case, the dynamics of Smoluchowski applies, since the number of aggregates change as the number of drops \cite{Urbina2006,Urbina2009}. Notice however, that the number of \emph{particles} decreases as a product of coalescence.

\subsubsection{Partial coalescence of flocculating drops (model M3)}

An initially monodisperse emulsion rapidly turns into a polydisperse system as a consequence of the flocculation and the coalescence of the drops. The turbidity of this system results from several contributions including primary (small) drops, aggregates of primary drops, bigger spherical drops resulting from the coalescence of small drops and ``mixed'' aggregates resulting from either the flocculation of big and small drops, or the partial coalescence of aggregates of small drops. If the number of ``mixed'' aggregates is negligible (see Appendix B):
\begin{equation}\label{TurbidityM3}
    \tau_{M3}=n_{1} \sigma_{1} + \sum_{k=2}^{k_{max}} n_{k,a} \sigma_{k,a} + \sum_{k=2}^{k_{max}} n_{k,s} \sigma_{k,s},
\end{equation}
where $k_{max}$ represents the maximum number of particles in an aggregate of size $k$. Subscripts $a$ and $s$ stand for aggregates and spherical drops, respectively.

If the large drops are originated during the process of flocculation of smaller drops, their number should be proportional to the values of $n_{k}$ proposed by Smoluchowski. The number of collisions predicted by Smoluchowski will be the same, but only a fraction $x_{k,a}$ of those collisions will produce aggregates of size $k$:
\begin{equation}\label{nka1}
    n_{k,a}=n_{k}x_{k,a}.
\end{equation}
If it is assumed that the fractions of aggregates ($x_{k,a}$) and drops ($x_{k,s}$)  are the same independent of the value of $k$, then:
\begin{equation}\label{nka2}
    n_{k,a}=n_{k}x_{k,a}= g n_{k},
\end{equation}
\begin{equation}\label{nks}
    n_{k,s}=n_{k}x_{k,s}= (1-g) n_{k}.
\end{equation}
Therefore,
\begin{equation}\label{TurbidityM3}
    \tau_{M3}=n_{1} \sigma_{1} + g \sum_{k=2}^{k_{max}} n_{k} \sigma_{k,a} + (1-g) \sum_{k=2}^{k_{max}} n_{k} \sigma_{k,s}.
\end{equation}

In the theory of Smoluchowski, an aggregate of $k$ particles contains $k$ primary particles. Therefore, its volume is equal to $k$ times the volume of a single primary particle. Hence, if the process of coalescence does not occur during the flocculation of drops but afterward, it will still affect the number of aggregates $n_{k}$ in the same way. The fraction $x_{k,s}$ results from both the instantaneous coalescences of the drops at contact or their ``delayed'' coalescences within an aggregate previously formed.

When coalescence does not occur and only the mechanism of aggregation operates, the fraction of flocs produced by aggregation is equal to one ($x_{k,a} = g = 1$). In this case, the third term on the right hand side of Eq. \ref{TurbidityM3} disappears, and Eq. \ref{TurbidezPuertas} is recovered.

Notice also that the summations of Eq. \ref{TurbidityM3} can be regrouped, giving rise to an average cross section for flocs and drops:

\begin{equation}\label{TurbidityM33}
    \tau_{M3}=n_{1} \sigma_{1} +  \sum_{k=2}^{k_{max}} n_{k} \; [g \; \sigma_{k,a} + (1-g) \: \sigma_{k,s}].
\end{equation}

\subsubsection{Modification of M1 to account for a finite coalescence time (model M4)}

Model M2 supposes that the coalescence time is negligible in comparison to the time required by the incident laser light to sample the structures of the system. If that is not the case, it might be possible to ``observe'' structures in which the flocculated drops are deformed but have not totally coalesced. This consideration is specially important if the deformation is such that (an approximately planar) film is formed between the drops. Small drops are not expected to deform significantly prior to coalescence. However, they have to form some sort of ``bridge'' between their internal phases before they merge into one another. In this case, the form factor of the transient structure would be intermediate between that of two flocculated particles and a bigger sphere. In order to test this possibility, we introduced an average cross-section for the transient structure equal to:
\begin{equation}\label{sigmatrans1}
    \sigma_{trans1}=\frac{\sigma_{k,a}\sigma_{k,s}}{(1-p) \sigma_{k,a} + p \, \sigma_{k,s}},
\end{equation}
where $p$ is a real number and $0\leq p \leq 1$. Notice that the use of equations of the type:
\begin{equation}\label{sigmatrans2}
    \sigma_{trans2}= p \, \sigma_{k,a} + (1-p) \, \sigma_{k,s},
\end{equation}
leads to the same expression of turbidity resulting from model M3.

Use of Eq. \ref{sigmatrans1} allows to express $\tau$ in the following form:
\begin{equation}\label{TurbidityM4}
    \tau_{M4}=n_{1} \sigma_{1}+ \sum_{k=2}^{k_{max}} n_{k} \sigma_{trans1}.
\end{equation}
This equation is equal to Eq. \ref{TurbidityM33} except for the expression of the cross section used in this case: $\sigma_{trans1}$. Thus, when $p=0$ only the spherical drops will contribute. In this case, the optical cross section $\sigma_{trans1}$ (Eq. \ref{sigmatrans1}) becomes equal to $\sigma_{k,s}$ and Eq. \ref{TurbidityM2} is recovered. This represents the case where only coalescence operates. Conversely, if only flocculation occurs $p=1$, $\sigma_{trans1} = \sigma_{k,a}$ and Eq. \ref{TurbidezPuertas} is recovered.

\section{Experimental}
\label{Experimental}

\subsection{Materials}
\label{Materials}
N-dodecane (Merck, 98\% purity) was purified by passing it twice through a column with alumina. Sodium dodecyl sulfate, SDS (Merck) was recrystallized from ethanol two times. NaCl (Merck, 99.5\% purity) and isopentanol (Scharlau Chemie S. A., 99\% purity) were used as received. The water used in all experiments was distilled and deionized (conductivity $< 1 \mu S cm^{-1}$ at 25 $^{\circ}$C) using a Simplicity purificator (Millipore, USA).

\subsection{Phase diagram [NaCl] vs. [SDS]}
\label{Phase diagram}

In order to select a starting system for nanoemulsion preparation, unidimensional scans of salt concentration (wt\% NaCl) were made at several surfactant concentrations (wt\% SDS). The weight-fraction of water was kept constant at $f_{w}=0.20$ ($\phi=0.84$). The isopentanol concentration was also maintained at 6.5 wt.\%. This alcohol was added in order to facilitate the transfer of the surfactant to the oil phase. Each system was prepared in vials, weighing each component to the final composition. The substances were added in the following order: water, SDS, isopentanol, dodecane and NaCl. The vials were subsequently gently shaken with a Varimix (Barnstead International, USA) for 2 hours in order to facilitate the contact between phases. Finally, the systems were placed in a thermostated bath at 25 $^{\circ}$C until equilibrium was reached, i.e., until complete phase separation was observed. The number of phases was evaluated visually. Anisotropic liquid crystalline phases were identified using polarized light.

\subsection{Nanoemulsion preparation}
\label{Nanoemulsion preparation}

\emph{Water/NaCl/isopentanol/SDS/dodecane} nanoemulsions were prepared using the phase inversion composition method \cite{Sole2006,Wang2008}. A $W_{t}+LC+O$ system was selected as a starting system. This system was suddenly diluted at constant stirring until a final weight-fraction of water of $f_{w}=0.95$ ($\phi=6.5 \times 10^{-2}$). This procedure allowed to obtain a \emph{concentrated mother} nanoemulsion. The \emph{final} nanoemulsions were obtained diluting the mother nanoemulsion with a suitable aqueous solution of SDS. During the dilution, the surfactant concentration and the number of drops per unit volume were adjusted to $8\times10^{-3}$ M and $2.0\times10^{11}$ drops/mL, respectively. The number density of the drops was estimated considering the mean radius of droplets obtained from photon correlation spectroscopy (PCS) measurements.

\subsection{Nanoemulsion characterization}
\label{Nanoemulsion Characterization}

The $\zeta$-potential of the drops was evaluated using a Zetasizer-Z (Malvern, UK). This equipment employs the Smoluchowski approximation \cite{Smoluchowski1903} which relates the mobility to $\zeta$-potential. The average droplet size and the droplet size distributions (DSD) of the nanoemulsions were evaluated by PCS using a Malvern Zetasizer 2000 (Malvern, UK).

The form of the DSD at a selected set of times was determined using a BI-200SM goniometer (Brookhaven Instruments, USA).

\subsection{Experimental determination of stability ratio ($W$)}
\label{Nanoemulsion stability}

In order to determine the stability ratio of a nanoemulsion, the change of the absorbance as a function of time at different electrolyte concentrations was measured. The optimal experimental conditions were obtained after several preliminary studies. These included: a) the adjustment of the velocity of addition of the salt solution in order to ensure the complete homogenization of the sample, and b) the determination of the optimal concentration of drops (from $dAbs/dt$ vs. $[Drops/mL]$ curves at different wavelengths) in order to avoid multiple scattering \cite{Maroto1997}.

A Genesys 5 spectrophotometer (Thermo Scientific, USA) was used, at a wavelength of $\lambda=$ 400 nm. At this lambda, the emulsion components do not absorb significantly, and the fulfillment of \emph{RGD} conditions is guaranteed. The methodology employed was very similar to the one previously used for studying the aggregation of latex suspensions \cite{Puertas1998,Rubio1994,Peula1997,Romero-Cano2001}: 2.4 mL of the nanoemulsion was placed into the cell of the spectrophotometer and its optical absorbance was measured. Following, 0.6 mL of the respective NaCl solution was quickly added in order to reach the desired electrolyte concentration. The salt concentration was changed between 380 and 600 mM. In all cases, the concentration of surfactant in the cell was kept constant at $8\times10^{-3} M$. Higher surfactant concentrations promote the formation of hydrated crystals of SDS above 600 mM NaCl.

The optical absorbance as a function of time was recorded continuously using a computer for a period of $30$ seconds. The measurements were repeated at least three times at each salt concentration. Since the aggregation process is very fast, the initial value of the absorbance after the addition of salt solution is unknown. An approximate value of the initial absorbance ($Abs([NaCl=0])$) was obtained, diluting 2.4 mL of the nanoemulsion with 0.6 mL of pure water. This was done previously to each experiment. From the value of this absorbance, the initial time of aggregation ($t_{0,exp}$) was estimated in the actual experiment.

\subsection{Evaluation of $W_{exp}$}

As a first approximation to $W$, the initial slope of the curves of $Abs$ vs. $t$ was considered to be proportional to the coagulation rate \cite{Romero-Cano2000,Puertas1997,Romero-Cano2001}. Notice that neglecting the difference in the radius of different preparations of the nanoemulsion, the cross-sections of the aggregates can be simplified in Eq. \ref{Constante-dimeros} giving in Eqs. \ref{W-turbidez} and \ref{W-turbidez2}. The value of $W_{exp}$ (Eq. \ref{W-turbidez}) was approximated by the quotient between the average initial slope obtained under fast aggregation conditions (between 420 and 600 mM of NaCl), and the one obtained at diluter electrolyte concentrations (380-410 mM NaCl).

\subsection{Evaluation of $W_{k_{11}}$}

Using Eq. \ref{Constante-dimeros} and the initial slopes of $Abs$ vs $t$ at each salt concentration, the rate for doublet constant formation ($k_{11}$) was calculated. The cross-sections, $\sigma_{1}$ and $\sigma_{2}$ were estimated using Eq. \ref{sigmak} for $k=1$ and $k=2$, respectively. These equations include the average radius ($R_{exp}$) of the emulsions. Since one mother nanoemulsion was prepared for each salt concentration, the values of $\sigma_{1}$ and $\sigma_{2}$ are slightly different for each preparation. Hence, the quotient $W_{k_{11}}=k_{11}^{fast}/k_{11}^{slow}$ (Eq. \ref{W-turbidez2}) is not necessarily equal to $W_{exp}$:
\begin{equation}
 W_{exp} \; \frac{\left(\frac{1}{2}\sigma_{2}-\sigma_{1}\right)^{slow}}{\left(\frac{1}{2}\sigma_{2}-\sigma_{1}\right)^{fast}} \,= \, W_{k_{11}}
\end{equation}

\subsection{Fitting of the turbidity curve}

The models M1, M2, M3 and M4 were programmed using a symbolic manipulation program (Mathematica 8.0.1.0). This program uses least-squares fittings in order to adjust the theoretical model of the turbidity to the experimental data. In all cases, the absorbance data was converted to turbidity using the relation: $\tau=230Abs$. The values of $k_{f}$, $t_{0}$, $g$, and $p$ are obtained from the fittings. The code of model M1 was tested using the original data from Ref. \cite{Puertas1997}, reproducing the same kinetic constant previously obtained.

The average radius ($R_{fit}$) of the drops is a parameter of the calculation, which in some cases was slightly varied in order to: a) maximize the quality of the fitting, and b) obtain a theoretical value of $t_{0,fit}$ close to the experimental one ($t_{0,exp}$).

Once the flocculation rates were obtained from the theoretical models, the stability ratios were calculated for each salt concentration.

\subsection{Polydispersity}

In order to appraise the effect of the initial polydispersity of the emulsions on the quality of the fittings of $k_{f}$, we recalculated the flocculation rate using the normalized frequency of the drop size distribution for three particle radii: $R_{min}=R_{fit}-\alpha_{sd} \, \sigma_{sd}$, and $R_{max}=R_{fit}+\alpha_{sd} \,\sigma_{sd}$, where $\sigma_{sd}$ is the standard deviation of the distribution and $\alpha_{sd}=1,2$. Using these radii, an approximate value of the turbidity can be calculated as:
\begin{equation}\label{Tauaprox}
 \tau_{aprox}=\tau(R_{min})f_{R_{min}}+\tau(R_{fit})f_{R_{fit}}+\tau(R_{max})f_{R_{max}},
\end{equation}
where $f_{R}$ stands for the frequency of the DSD corresponding to radius $R$ ($f_{R_{min}}+f_{R_{fit}}+f_{R_{max}}=1$). The effect of $R_{min}$ and $R_{max}$ was observed, comparing the behavior of $\tau_{aprox}$ with the one of $\tau({R_{fit}})$ for models M1 and M3 at [NaCl]$=$390 mM and 475 mM. The difference between $R_{exp}$ and $R_{fit}$ for 390 and 475 mM NaCl is 14 and 17 nm, respectively.

\section{Results}
\label{Results}

\begin{figure}[htbp]
\centering
    \includegraphics[width=9cm]{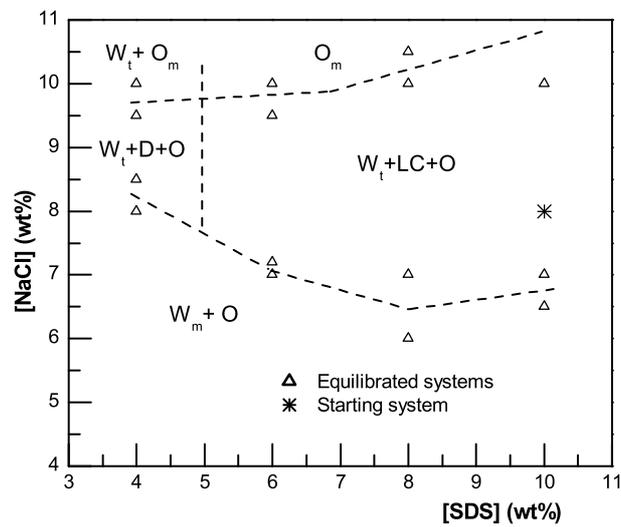}\\
  \caption{[NaCl] vs. [SDS] phase diagram of the \emph{water/NaCl/isopentanol/SDS/dodecane} system at $6.5$ \%wt isopentanol and $f_{w}=0.20$.}\label{DiagramaNaClSDSfinal}
\end{figure}

Fig. \ref{DiagramaNaClSDSfinal} shows the phase behavior of the systems as a function of surfactant and salt concentrations. The dash lines indicate the approximate limits of different phase behaviors. The three-phase region is located in the center of the diagram. This region is divided into two zones: a) a subregion containing water ($W_{t}$), a bicontinuous microemulsion $(D)$ and an oil phase $(O)$, which is formed at a low surfactant concentration (4 wt.\% SDS); and b) a second subregion composed of water ($W_{t}$), liquid crystal $(LC)$ and an oil phase $(O)$, formed at high surfactant concentrations ($>6$ wt.\% SDS) \cite{Ekwall1975}. Below the three-phase region, a two-phase region consisting in a direct micellar solution $(W_{m})$ and an excess oil phase $(O)$ is observed. Here, the surfactant is preferentially dissolved into the aqueous phase and the formation of $o/w$ emulsions is favored. Above three-phase zone, a two-phase region exists, consisting of excess of water $(W_{t})$ in equilibrium with a reverse micellar solution $(O_{m})$. When the surfactant concentration is higher than $6$ wt.\%, a one-phase region appears, $O_{m}$.

As Fig. \ref{DiagramaNaClSDSfinal} indicates, it is possible to promote a phase transition changing the salt concentration. During a transition involving three phases, the system shows low values of interfacial tension, which favors the formation of drops of nanometer size. According to previous experimental works \cite{Izquierdo2002,Izquierdo2004,Morales2006,Gutierrez2008,Forgiarini2001}, the presence of liquid crystals favors the formation of stable nanoemulsions. Hence, \emph{Water/NaCl/isopentanol/SDS/dodecane} nanoemulsions were prepared starting from a $W_{t}+LC+O$ system.

\begin{figure}[htbp]
\centering
    \includegraphics[width=9cm]{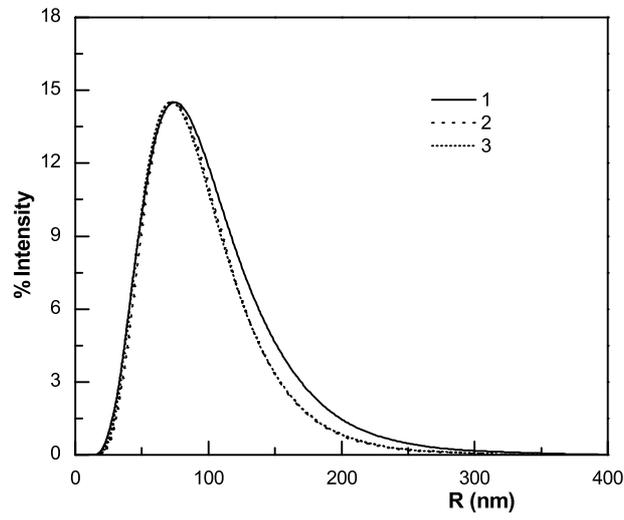}\\
  \caption{Droplet size distributions of three independent nanoemulsions synthesized from three different mother emulsions measured with a Malvern Zetasizer 2000.}\label{DSD2}
\end{figure}

The size distribution of a typical nanoemulsion prepared by the phase inversion composition method \cite{Sole2006,Wang2008} is shown in Fig. \ref{DSD2}. It has a log-normal distribution with an average radius around $73$ nm. The physical characteristics of the dispersion reported in Table \ref{Nanocharacteristics} correspond to the average of ten nanoemulsions prepared independently. The composition of the starting system, the concentrated mother emulsion and the final diluted nanoemulsion are shown in Table 1.

\begin{figure}[htbp]
\centering
    \includegraphics[width=9cm]{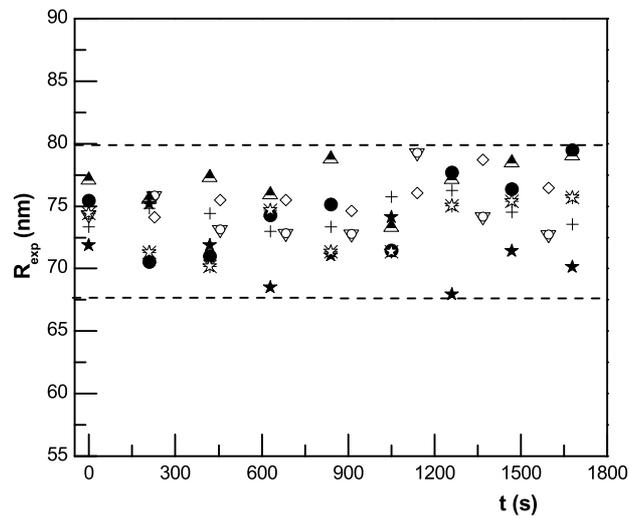}\\
  \caption{Evolution of $R_{exp}$ as a function of time for several nanoemulsions.}\label{Radiostodos}
\end{figure}

Previous to the turbidity measurements, the evolution of the average droplet radius ($R_{exp}$) of several nanoemulsions as a function of time was studied (see Fig. \ref{Radiostodos}). In the case of emulsions, it is uncertain whether the fluctuations observed in Fig. \ref{Radiostodos} correspond to the statistical error of the measurements or they reflect a true variation of the average radius. In any event, it is clear that the average radius does not change considerably within the first $30$ minutes. This time it is substantially longer than the one required for the evaluation of the turbidity at each ionic strength. Nevertheless, one \emph{mother} nanoemulsion was synthesized for each salt concentration.

\subsection{Variation of turbidity as a function of time}
\label{Results-W}

\begin{figure}[htbp]
\centering
   \includegraphics[width=9cm]{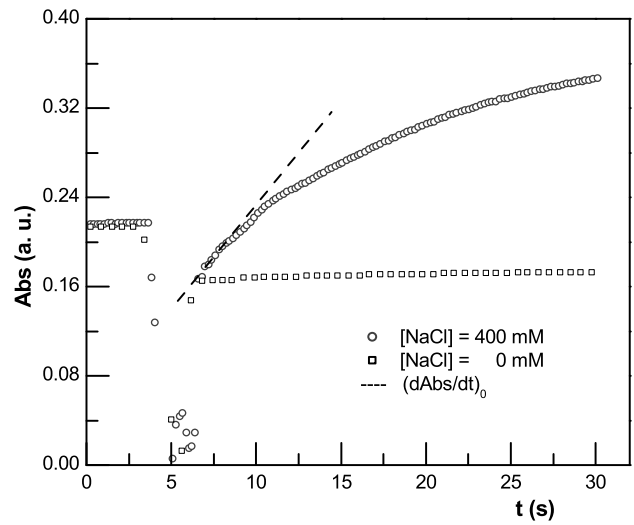}\\
  \caption{Typical curve of $Abs$ vs. $t$.} \label{Absvstiempo2}
  \end{figure}

Fig. \ref{Absvstiempo2} shows the typical curve of $Abs$ vs. $t$ of an aggregating colloidal dispersion. Prior to addition of the salt solution, the absorbance is constant. As soon as the 0.6 mL of the NaCl solution is added, the absorbance decreases due to the dilution of the emulsion. Then, it increases indicating the aggregation of the drops. As can be observed in Fig. \ref{Absvstiempo2}, there is a non-monotonous variation between 4 and 6 seconds. During this time, two events occur: a) the homogenization of the sample due to the addition of salt solution, and b) the undesired input of external light that enters the measuring cell when the compartment of the instrument is opened in order to add the salt solution. The process of flocculation begins at time $t_{0,exp}$ which occurs during this transient period. This value is used as a reference for the theoretical value, $t_{0,fit}$, obtained from the fitting. The comparison of these two values is an additional test of consistency of the theoretical models.

The relevant variation of $Abs$ vs. $t$ begins as soon as the monotonous curve is observed. At this point, the initial slope of the curve can be calculated.

\begin{figure}[htbp]
\centering
   \includegraphics[width=9cm]{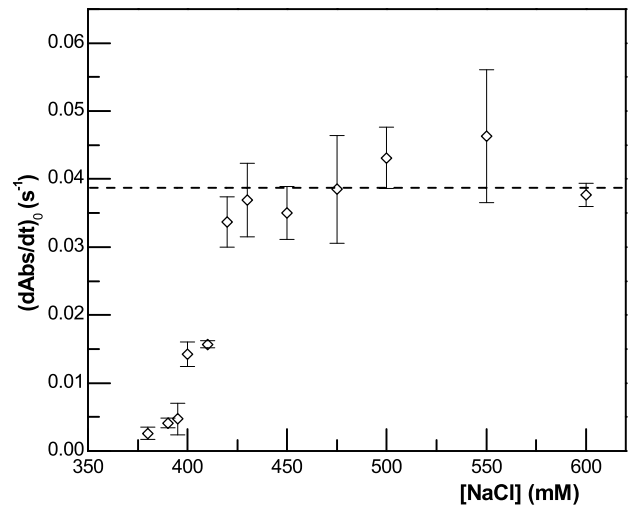}\\
  \caption{Aggregation kinetics of the nanoemulsion.}\label{PendienteSal}
\end{figure}

Fig. \ref{PendienteSal} shows the initial slope of $Abs$ vs. $t$ as a function of the salt concentration. Between 380 and 420 mM NaCl, the emulsion is slowly destabilized. In suspensions, this regime is associated to Reaction Limited Cluster Aggregation (RLCA). Above 430 mM NaCl, the emulsion destabilizes very fast, showing values of $dAbs/dt$ which are similar within the error bars (Appendix C). This regime corresponds -in the case of suspensions- to Diffusion Limited Cluster Aggregation (DLCA).

\begin{figure}[htbp]
\centering
   \includegraphics[width=16cm]{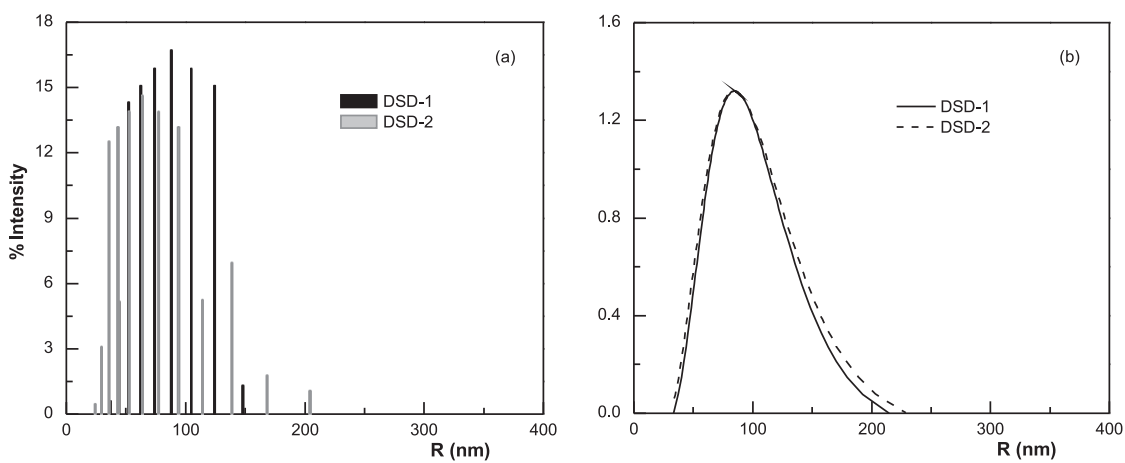}\\
  \caption{Results of two measurements of the initial droplet size distribution of the same nanoemulsion using a BI-200SM Goniometer: (a) Contin, (b) Cumulants analysis.}\label{DSDInicial2}
\end{figure}

\begin{figure}[htbp]
\centering
   \includegraphics[width=16cm]{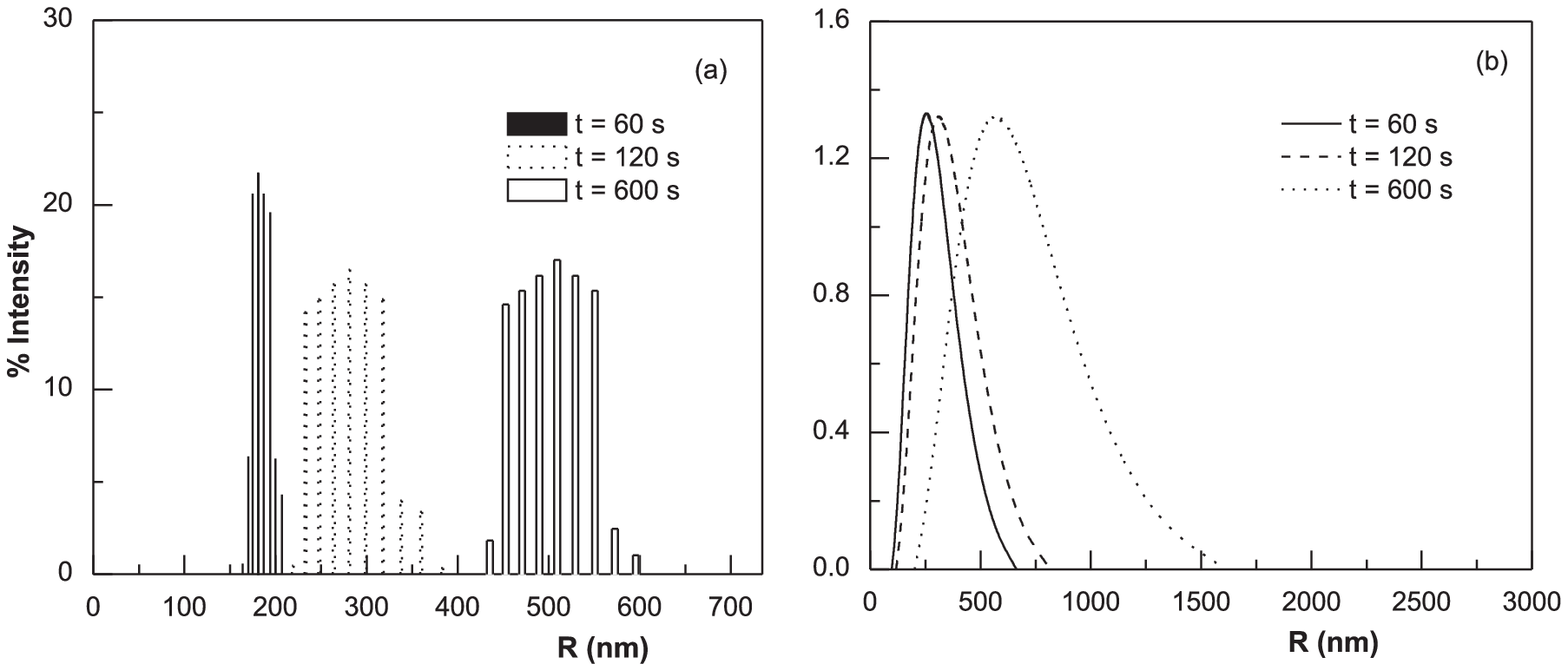}\\
  \caption{Temporal evolution of the droplet size distributions for a nanoemulsion with 380 mM NaCl using a BI-200SM Goniometer: (a) Contin, (b) Cumulants analysis.}\label{DSD380final}
\end{figure}

\begin{figure}[htbp]
\centering
   \includegraphics[width=16cm]{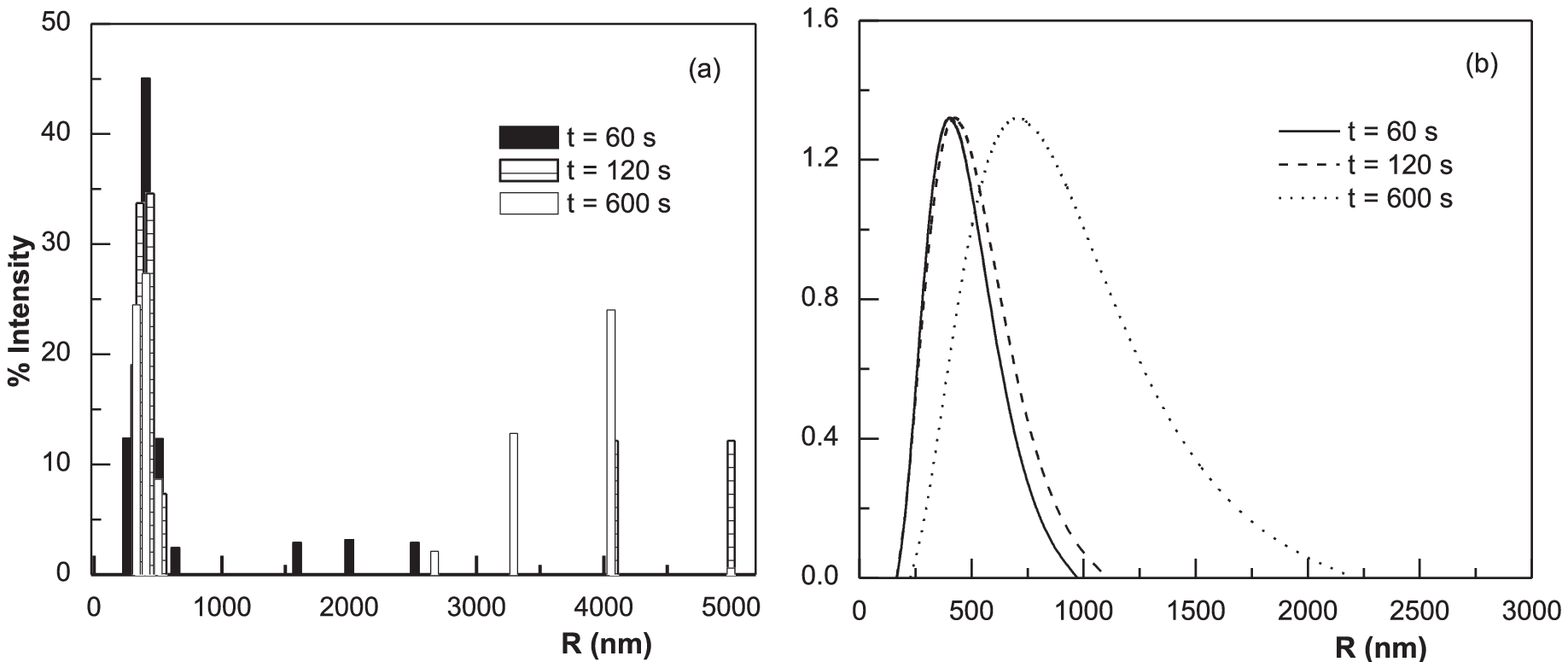}\\
  \caption{Temporal evolution of the droplet size distributions for a nanoemulsion with 600 mM NaCl using a BI-200SM Goniometer: (a) Contin, (b) Cumulants analysis.}\label{DSD600final}
\end{figure}

\begin{figure}[htbp]
\centering
   \includegraphics[width=16cm]{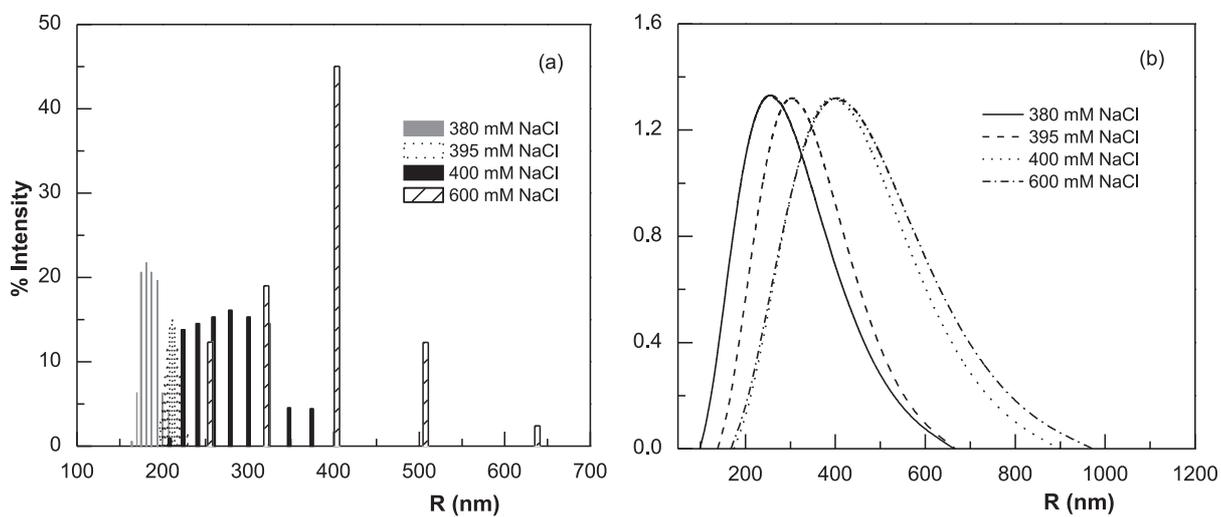}\\
  \caption{Droplet size distributions after 1 min of addition of different concentrations of salt using a BI-200SM Goniometer: (a) Contin, (b) Cumulants analysis.}\label{DSDSalesfinal}
\end{figure}

Fig. \ref{DSDInicial2} shows the results of two measurements of the initial droplet size distribution of the same nanoemulsion using the algorithms of Contin \cite{CONTINoriginal} and Cumulants analyses \cite{CUMULANTS} implemented in the Brookhaven BI-200SM Goniometer. Fig. \ref{DSD380final} illustrates the evolution of the DSD for an ``RLCA'' system ([NaCl] = 380 mM). The average radius and the width of the distribution increase monotonically as a function of time due to the aggregation of the drops. The width of DSD enlarges considerably after 600 s. At high ionic strength (Fig. \ref{DSD600final}, [NaCl] = 600 mM), the polydispersity of the system increases much more pronouncedly due to the coalescence of the drops (notice the scale of the abscissa). The distribution turns bimodal after 60 s, with peaks around 400 and 2000 nm according to the Contin analyses. Between these two extreme cases, a certain degree of stabilization is exhibited (Fig. \ref{PendienteSal}). Notice the difference between the DSD corresponding to 395 and 400 mM (Fig. \ref{DSDSalesfinal}). This is the coalescence threshold suggested by models M3 and M4. According to the simulations \cite{Urbina2005a}, the progressive decrease of interfacial area due to the coalescence of the drops and the dynamic adsorption equilibrium of the surfactant to the available interfaces generate a progressive increase of the repulsive forces between the drops as a function of time. If the drops remain non-deformable, the total surfactant concentration and the ionic strength of the solution will determine if complete phase separation occurs or a lower value of the total interfacial area of the emulsion can be stabilized.

\subsection{Fitting of the turbidity curve with theoretical models}

\begin{figure}[htbp]
\centering
  \includegraphics[width=9cm]{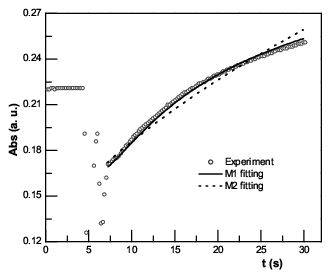}\\
  \caption{Comparison of the theoretical curves of absorbance predicted by models M1 and M2 with the experimental measurements of slow aggregation ($[NaCl]=390$ mM).}\label{wf2-largerspheres}
\end{figure}

\begin{figure}[htbp]
\centering
  \includegraphics[width=9cm]{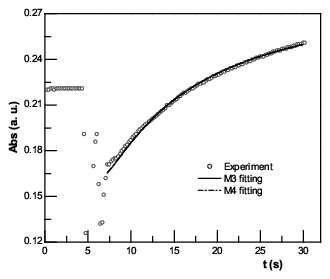}\\
  \caption{Comparison of the theoretical curves of absorbance predicted by models M3 and M4 with the experimental measurements of slow aggregation ($[NaCl]=390$ mM).}\label{wf2-mixcombine}
\end{figure}

\begin{figure}[htbp]
\centering
  \includegraphics[width=9cm]{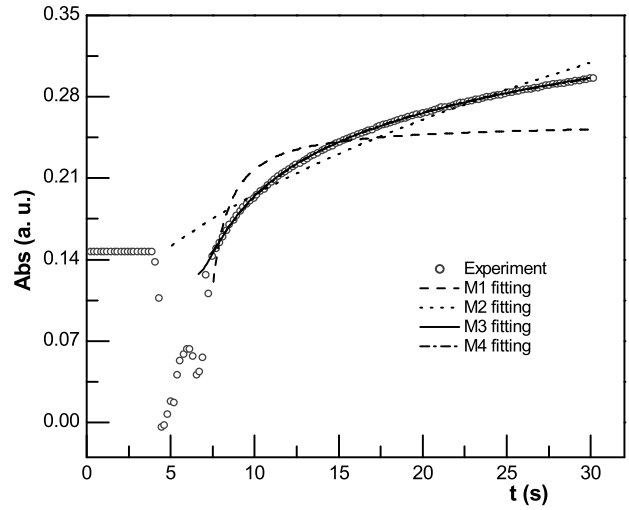}\\
  \caption{Comparison of theory and experiment for the case of fast aggregation ($[NaCl]=600$ mM).}\label{600todos}
\end{figure}

Table \ref{k11table} shows the kinetic constants obtained for doublet formation ($k_{11}$). Figs. \ref{wf2-largerspheres}, \ref{wf2-mixcombine} and \ref{600todos} show a comparison of the theoretical models for $\tau$ with the experimental measurements for the cases of slow (Figs. \ref{wf2-largerspheres} and \ref{wf2-mixcombine}) and fast aggregation (Fig. \ref{600todos}).

The data in Fig. \ref{wf2-largerspheres} correspond to $[NaCl]=390$ mM. The theoretical models were calculated using $k_{max}=100$ (Appendix D). While model M1 reproduces the data quite well, the model of spheres (M2) roughly follows the experimental trend. Fig. \ref{wf2-mixcombine} compares the same experimental data to the predictions of models M3 and M4. Both models fit the data satisfactorily. However, they showed a slight deviation at the initial times. Notice that the average aggregation time $t_{ave}=t_{1/2}=1/n_{0}k_{f}$ is of the order of $9.3$ s. Hence, the models are able to reproduce at least 3 $t_{1/2}$.

In the case of fast aggregation (Fig. \ref{600todos}), $[NaCl]=600$ mM, models M1 and M2 fail to reproduce the experimental data completely, while the models M3 and M4 still hold. It is impossible to distinguish between M3 and M4 in Fig. \ref{600todos} since both fittings are of similar quality.

\begin{figure}[htbp]
\centering
  \includegraphics[width=9cm]{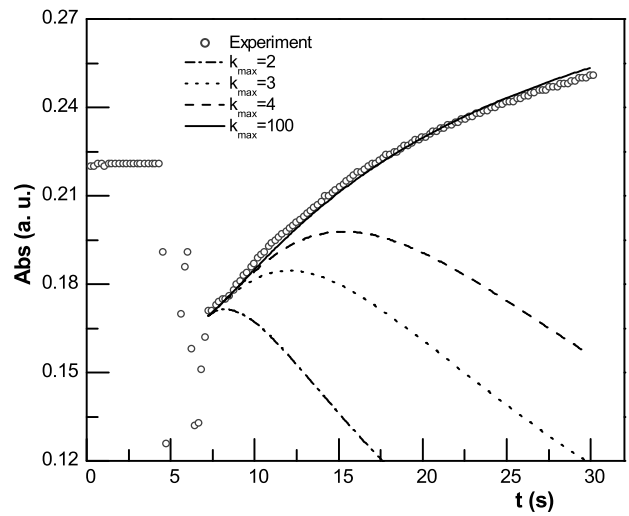}\\
  \caption{Predictions of model M1 for $k_{max}=2, 3, 4$ and $100$ ($[NaCl]=390$ mM).}\label{wf2-todos}
\end{figure}

For completeness, Fig. \ref{wf2-todos} shows the turbidity curves calculated with model M1 for $k_{max}=2,3,4$ and $100$, even for this 30-seconds evolution. As expected a large value of $k_{max}$ is necessary in order to reproduce the experimental curve completely, due to the high rate of the reaction.

\begin{figure}[htbp]
\centering
    \includegraphics[width=9cm]{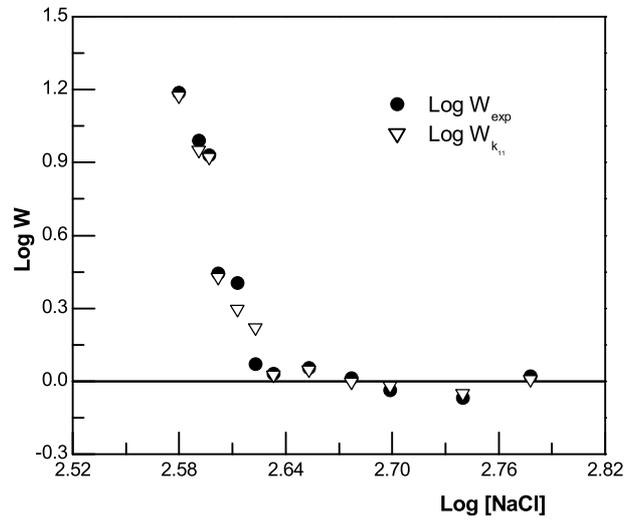}\\
  \caption{Comparison between $W_{k_{11}}$ and $W_{exp}$.}\label{LogWexpW11final}
\end{figure}

We move now to comparing the stability ratios. $W_{exp}$ and $W_{k_{11}}$ were obtained using Eq. \ref{W-turbidez} and Eq. \ref{W-turbidez2}, respectively. A plot of $W_{exp}$ vs. $[NaCl]$ clearly shows the regions corresponding to the slow and fast aggregation (see Fig. \ref{LogWexpW11final}). Using the kinetic constants of Table \ref{k11table}, the values of $W_{k_{11}}$ were calculated and compared with the experimental ones ($W_{exp}$). There is a very good agreement between both determinations.

Tables \ref{kM1}, \ref{kM3} and \ref{kM4} show the kinetic rates of flocculation obtained from the fitting of models M1, M3 and M4. In Table \ref{kM1}, the kinetic constants between 380 and 395 mM NaCl result from an appropriate fitting of the data by the theoretical model (M1). Above 400 mM NaCl, the model does not adjust to the experimental trend. Consequently, the values of the error bars of the kinetic constants are higher than the average value. Note that the particle radius, that has been chosen as a fitting parameter (Tables \ref{kM3} and \ref{kM4}) fluctuates around the value measured with PCS (Fig. \ref{Radiostodos}), showing that it can be taken as a robust measure, despite the inherent polydispersity of the system. Interestingly, this value of the particle radius is obtained from a fitting of the whole turbidity curve, i.e. where large aggregates are also included, and not only single particle data.

\begin{figure}[htbp]
\centering
    \includegraphics[width=9cm]{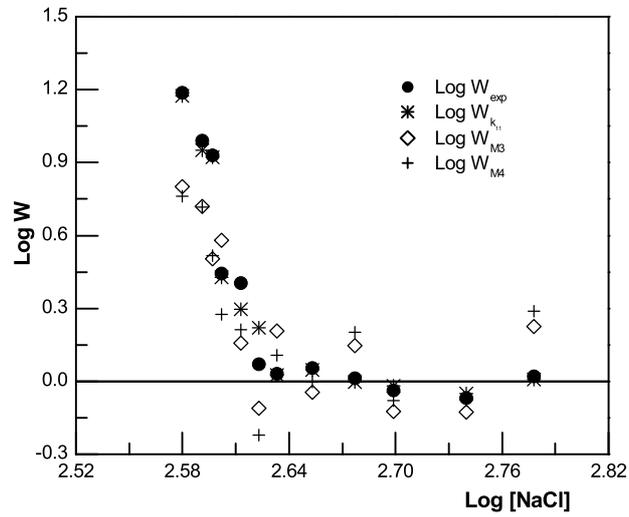}\\
  \caption{Comparison between theoretical and experimental values of $W$.}\label{LogtodoslosWfinal}
\end{figure}

The values of $W$ were determined from the data of Tables \ref{kM3} and \ref{kM4}. Fig. \ref{LogtodoslosWfinal} shows a comparison between the theoretical and experimental values of the stability ratio. The fast flocculation rates required to calculate $W_{k_{11}}$, $W_{M3}$ and $W_{M4}$ were approximated by the average of the values of $k_{f}$ obtained between $420$ and $600$ mM NaCl ($k_{11}^{fast}=$3.2 $\times 10^{-18}$ $m^{3} s^{-1}$, $k_{M3}^{fast}=$1.3 $\times 10^{-18}$ $m^{3} s^{-1}$, and $k_{M4}^{fast}=$ 1.2 $\times 10^{-18}$ $m^{3} s^{-1}$). The agreement is satisfactory considering the complexity of the emulsion system. In the case of emulsions, the turbidity is associated with the change in concentration of the droplets, which is due to the phenomena of flocculation, coalescence and Ostwald ripening. While in the case of solid suspensions, the initial slope of turbidity curve is only due to aggregation, and more specifically, to the formation of doublets \cite{Lichtenbelt1973}.

The values of the critical coagulation concentration (CCC) of the nanoemulsion predicted from the experimental data and the theoretical fittings (Fig. \ref{LogtodoslosWfinal}) are shown in Table \ref{CCC}.

\subsection{Effect of the initial polydispersity on the evaluation of $k_{f}$}

\begin{figure}[htbp]
\centering
    \includegraphics[width=9cm]{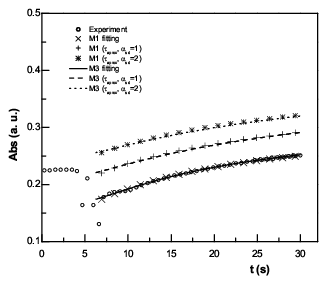}\\
  \caption{Comparison of the behavior of $\tau_{aprox}$ with $\tau (R_{fit})$ for models M1 and M3 at $[NaCl]=390$ mM. (The number of data points of the experimental curve was reduced in order to highlight the quality of the fittings).}\label{Poly390}
\end{figure}

\begin{figure}[htbp]
\centering
    \includegraphics[width=9cm]{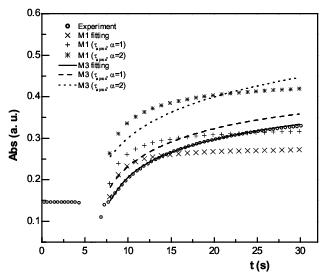}\\
  \caption{Comparison of the behavior of $\tau_{aprox}$ with $\tau (R_{fit})$ for models M1 and M3 at $[NaCl]=475$ mM. (The number of data points of the experimental curve was reduced in order to highlight the quality of the fittings).}\label{Poly475}
\end{figure}

Figs. \ref{Poly390} and \ref{Poly475} illustrate the effect of the initial polydispersity on the calculation of $k_{f}$. As the polydispersity increases, the theoretical value of the turbidity deviates from the experimental points. This happens for both models M1 and M3 at both salinities tested (390 and 475 mM NaCl). However, the deviations are larger in the case of model M1 at 475 mM since it does not adjust the experimental data. Since model M3 converges to model M1 when only aggregation occurs, the deviations of these models are very similar for [NaCl]$=$ 390 mM. Similar results are obtained using model M4 instead of M3. In both cases (M3 and M4) an increase in the polidispersity of the system provokes a displacement of the turbidity curve, without changing its curvature. This implies that a smaller radius should be used to reproduce the experimental data, but leaves our main conclusion unchanged, namely, the interplay of coagulation and coalescence have to be considered to describe correctly the destabilization of nanoemulsions, such as is done in our models M3 and M4.

\section{Discussion}

As shown in Fig. \ref{PendienteSal}, a dodecane-in-water nanoemulsions stabilized with SDS shows the typical behavior of charged latex suspensions with respect to the salt concentration. Below the critical coagulation concentration ($CCC=411$ mM NaCl) the system is kinetically stabilized with respect to flocculation. The stability decreases progressively with the increase of the ionic strength. Above the \emph{CCC}, the flocculation of the particles occurs very fast, and the aggregation rate reaches a plateau, similar to the one exhibited by solid particles under DLCA conditions.

It is important to remark that each point in Fig. \ref{PendienteSal} corresponds to the average at least three different measurements. During the first $30$ minutes, the emulsion is fairly stable: its average radius changes less than $10\%$. After this time, the average radius changes considerably even in the absence of an appreciable concentration of salt. This instability probably occurs as a consequence of Ostwald ripening \cite{Urbina2009a,Izquierdo2002}, which is favored due to the high polydispersity of the emulsions and the appreciable solubility of the dodecane in water.

In the case of suspensions, the addition of salt screens the surface charges of the particles promoting their flocculation. In the case of emulsions, the charge of the surfactant adsorbed to the oil/water interface is also screened, but further surfactant adsorption is also promoted. Hence, the stability of the emulsion results from a balance between these two phenomena. Once the salt concentration is high enough to produce a maximum surfactant adsorption, subsequent addition of salt will only generate the progressive screening of the surface charges as predicted by the theory of DLVO.

At [NaCl] $< CCC$ (see Table \ref{CCC}), the original drops (primary particles) are expected to aggregate following the dynamics proposed by Smoluchowski \cite{Urbina2006,Urbina2009}. Thus, Eqs. \ref{Smoluchowski}, \ref{W-experimental} and \ref{nkPuertas} should be observed. The initial slope of the curve of $Abs$ vs. $t$ at each salt concentration can be used along with Eq. \ref{Constante-dimeros} to obtain the rate of doublet formation ($k_{11}$). Table \ref{k11table} shows a minimum value of $2.1 \times 10^{-19}$ $m^{3} s^{-1}$. The highest value obtained corresponds to $3.8 \times 10^{-18}$ $m^{3} s^{-1}$. This value is very similar to the one reported by Puertas and de las Nieves \cite{Puertas1997} for the case of latex particles (MP3, R\,=\,92.5 nm). As shown in Fig. \ref{LogWexpW11final}, the stability ratios $W_{exp}$ and $W_{k_{11}}$ are very similar.

Model M1 represents the effect of the aggregation of the drops on the turbidity. This model fits the data for $380 \leq [NaCl] \leq 395$ mM NaCl within the slow aggregation regime. This is illustrated in Fig. \ref{wf2-largerspheres} for the case of $[NaCl]=$ 390 mM. As shown in Fig. \ref{wf2-todos}, it is necessary to include a large number of aggregates (large $k$) in order to reproduce the absorbance of the emulsion for $t =$ 30 s.

If coalescence occurs, the rate constants determined through the turbidity measurements also sample this process. The effect of coalescence should be more pronounced as the salt concentration increases. In the extreme case in which only attractive forces prevail and coalescence occurs instantaneously as the drops collide, the collisions lead to larger drops. The number of drops should change as predicted by Smoluchowski for the number of aggregates (Eq. \ref{nkPuertas}) with the sole difference that the aggregates of size $k$ are now bigger drops with the equivalent volume of $k$ primary drops (Eq. \ref{ak}). Model M2 represents this case. Curiously, it completely fails to reproduce the data corresponding to fast aggregation.

Model M3 represents the situation in which some collisions between the drops lead to flocculation while others lead to coalescence. The turbidity of an emulsion results from the contributions of conventional aggregates, bigger drops and mixed aggregates (see Appendix B). If the number of mixed aggregates is small, the turbidity can be expressed as a combination of large drops and conventional aggregates (Eqs. \ref{nka2}, \ref{nks} and \ref{TurbidityM3}). When the salinity is low and only aggregation occurs, $g \longrightarrow$1.0. If $g$ is assumed to be equal to 1.0 in Eq. \ref{TurbidityM3}, the equation of the turbidity for M1 is recovered. Hence, model M3 is capable of reproducing the phenomenon of aggregation in the absence of coalescence. As can be inferred from Figs. \ref{wf2-mixcombine} and \ref{600todos}, model M3 adjusts the turbidity data corresponding to the slow and fast aggregation. Unfortunately, the fraction of aggregates represented by coefficient ``$g$'', surpassed the value of 1.0 for 380 $\leq [NaCl] \leq$ 390 mM. This indicates that the number of mixed aggregates cannot be neglected, and that their average effect is being accommodated within the contributions of the conventional aggregates and the bigger drops. This can only be possible if the form factor of the mixed aggregates is somewhere in-between these two contributions, otherwise the fitting of the experimental data will fail if mixed aggregates exist. In this regard, it should be noticed that Eq. \ref{TurbidityM3} can be cast into the form of Eq. \ref{TurbidityM33}. This last equation is similar to Eq. \ref{TurbidityM4} if the expression inside the brackets is regarded as an average cross section. The cases in which $g < 1$ correspond to fast aggregation (see Table \ref{kM3}). In these systems, between $60\%$ and $80\%$ of the primary drops are assembled into conventional aggregates, and the rest correspond to bigger separate drops resulting from coalescence.

In order to deduce Eq. \ref{TurbidityM3} analytically, it was assumed that the fraction of aggregates ($x_{k,a}$) and the drops ($x_{k,s}$) is the same for each particle size ($x_{k,a} = x_{a}$, $x_{k,s} = x_{s}$). This requirement is rather difficult to attain in practice. On the one hand, it can happen for example, if coalescence occurs during binary collisions only, and intra-cluster coalescence does not occur. Hence, only a fraction of the collisions will lead to larger aggregates, and the drops will grow by stepwise addition of primary particles. On the other hand, mixed doublets are commonly observed in macroemulsions \cite{Holt1997,vandenTempel1953}. This is an indication that either intracluster coalescence or hetero-aggregation occurs for radii of several microns.

It is also possible that the fraction of aggregates and spheres is similar but not equal at each particle size ($x_{k,a} \sim x_{a}, x_{k,s} \sim x_{s}$). In this case, the condition $0 \leq g \leq 1$ will not be fulfilled, but Eq. \ref{TurbidityM3} will still be approximately valid. However, in the general case mixed aggregates cannot be disregarded under RLCA conditions. We attempted to solve Eq. \ref{Turbidity3-0} for [NaCl]$=$395 mM, using Eqs. \ref{sigmatrans1} or \ref{sigmatrans2} as approximate cross sections. Only the use of Eq. \ref{sigmatrans2} fitted the data. But even in this case, $x_{a} < 0$, and $x_{m} < 0$, although $x_{a} + x_{s} + x_{m} = 1.02$. Therefore, it is clear that the reason why Eq. \ref{Turbidity3-0} is successful in reproducing the experimental trend deserves further research.

Model M4 considers a different aspect of the problem. If the process of coalescence is slow enough, the form factor observed by the incoming light of the spectrophotometer might correspond to a transient structure, in-between a doublet and a bigger spherical drop. If the cross-section of the transient structure is approximated as a sum of spherical and conventional aggregate contributions, the expression of the turbidity corresponding to model M3 is recovered. Consequently, we tested the analytical form given by Eq. \ref{sigmatrans1}. Unfortunately, the results of model M4 fitted all experimental data with a similar accuracy than model M3. Hence, the experimental data does not allow to discriminate the physical origin of the fittings. Since the use of $\sigma_{trans2}$ (Eq. \ref{sigmatrans2}) in model M4 (Eq. \ref{TurbidityM4}) reproduces the equation of the turbidity of model M3 (Eq. \ref{TurbidityM3}), these theoretical approaches rather represent two different interpolations of the cross-section of mixed aggregates occurring in emulsions (see Eq. \ref{TurbidityM33}). It can be affirmed that only those models which consider the process of partial coalescence can fit the experimental data corresponding to a high ionic strength. Moreover, according to the results of the previous paragraph, Eq. \ref{sigmatrans2} appears to be a better approximation for the optical cross section of the aggregates than Eq. \ref{sigmatrans1}. Despite these limitations and the ones previously mentioned regarding the effect of polydispersity, the mixed flocculation-coalescence rates observed at high ionic strength (Tables \ref{kM3} and \ref{kM4}) are of the same order of magnitude as the one predicted by Emulsion Stability Simulations (\emph{ESS}) for the case of very dilute emulsions in the absence of a repulsive force is $k_{f} = k_{FC} = 6.4 \times 10^{-18}$ $m^{3} s^{-1}$ \cite{Urbina2004}. Moreover, the $k_{f}^{fast}$ obtained are in most cases, half the value of $k_{11}$ as predicted by the theory of Smoluchowski.

Fig. \ref{LogtodoslosWfinal} shows a comparison of all stability ratios calculated in this work. The ones corresponding to models M3 and M4 were calculated using the flocculation rates of Tables \ref{kM3} and \ref{kM4}. In our opinion the agreement is satisfactory.

Finally, it is necessary to discuss the effect of polydispersity, given our emulsions are rather polydisperse (CV = 35\%). The influence of polydispersity has been investigated by many authors (Mullholland \cite{Mulholland1977}, Friedlander and Wang \cite{Friedlander1966,Wang1967}, Cohen and Vaughan \cite{Cohen1971}, and Lee \cite{Lee1983}). The continuous distribution of particle sizes is a non-linear integro-differential equation whose kernel depends on the size of the colliding aggregates. Meesters and Ernst \cite{Meesters1987} demonstrated that the use of a constant kernel in this equation leads to an exponential frequency distribution of particle sizes. Smoluchowski result approaches this exponential distribution at long times. If a homogeneous kernel is used, and the degree of homogeneity is greater than zero, bell-shaped distributions are obtained \cite{Elimelech1995}. In this case, collisions between large and small aggregates occur more rapidly than in the case of a constant kernel. This extra depletion of smaller aggregates leads to bell-shaped particle size distributions.

In regard to polydispersity the work of Lee \cite{Lee1983} is especially relevant to this investigation. This author assumed that the initial and subsequent drop-size distributions could be represented by a log-normal function. In this case, an expression for the number of aggregates \emph{produced by flocculation} -similar to the one of Smoluchowski- can be deduced:

\begin{equation}\label{Lee1}
    n_{a}=\frac{n_{0}}{(1+C_{p} \, k_{f}n_{0}t)},
\end{equation}
where:
\begin{equation}\label{Lee2}
   C_{p}= 1 + \exp (ln^{2} (\overline{\sigma}_{0})),
\end{equation}
and $\overline{\sigma}_{0}=\overline{\sigma} \,(t=0)$.

Coefficient $C_{p}$ corrects the flocculation kernel for polydispersity. According to Lee \cite{Lee1983} the size distribution approaches that having a geometric standard deviation of $\overline{\sigma}_{\infty}=$ 1.32 regardless of the value of $\overline{\sigma}_{0}$. The difference between the predictions of Eq. \ref{Smoluchowski} and the ones of Eq. \ref{Lee1} for an initially monodisperse system is less than 4\%:  $C_{p} =$ 1.08 \cite{Lee1983} ($C_{p} =$ 1.13  according to Friedlander and Wang \cite{Friedlander1966}). However, if the initial distribution is polydisperse, the difference between both equations becomes much larger, and increases with the value of $\overline{\sigma}_{0}$. Even in this case the asymptotic limit is approached, but at considerably longer times. In any event, it should be noticed that coefficient $C_{p}$ multiplies $k_{f}$. Hence, a fit of the data to Eq. \ref{Smoluchowski} leads to an uncertainty in the value of $k_{f}$ which increases with the initial polydispersity of the system.

It is clear from Figs. \ref{Poly390} and \ref{Poly475} that the increase of the polydispersity of the initial (mother) emulsions decreases the quality of the fittings. However, model M3 is less susceptible to this variable than model M1. This is probably due to its additional fitting parameter ($g$). In any case, the introduction of polydispersity does not change the curvature of the absorbance curve, supporting our main results that models M1 and M2 do not describe the process correctly at high salinity, whereas M3 and M4 fit the data correctly.

\section{Conclusions}
\label{Conclusions}

The stability of a dodecane/water nanoemulsion stabilized with SDS was evaluated using turbidity measurements. For this purpose, the variation of the turbidity as a function of time was expressed in terms of the flocculation rate of the dispersion. Four approximate theoretical models were postulated in order to represent the contribution of the aggregates of the emulsion to the turbidity. The complete set of experimental data was only reproduced by those models which take into account the existence of the conventional aggregates as well as bigger spherical drops resulting from coalescence. The adjustment of the experimental turbidity to the corresponding theoretical expressions allowed the determining of the value of the mixed flocculation-coalescence rate. The stability ratios ($W$) evaluated using these rates are consistent with those determined experimentally from the initial slope of the absorbance as a function of time.

As previously pointed out, the complexity of the emulsions results from the coupling of several destabilization mechanisms. In this report, a "window" of compositions where the processes of flocculation and coalescence outweigh significantly other destabilization mechanisms is found. It is important to remark that the mixed flocculation-coalescence rates shown in Tables \ref{kM3} and \ref{kM4} do not correspond to the original nanoemulsion synthesized. They only appraise the effect of the referred processes in the presence of a considerable amount of salt. Several technical difficulties appear when the salt concentration is lowered much further ($< 300$ mM). Hence, it is not generally possible to evaluate the flocculation rate of the original emulsion. This restriction also applies in the case of suspensions.

\section{Acknowledgements}
\label{Acknowledgements}

This work was supported by the Agencia Espa\~nola de Cooperaci\'on Internacional para el Desarrollo (AECID) under project $A/024004/09$. K. Rahn-Chique and C. Rojas gratefully acknowledge Dr. Aileen Lozs\'an for her invaluable support and helpful discussions. A.M. Puertas and M.S. Romero-Cano acknowledge financial support from the Ministerio de Ciencia y Tecnolog\'ia and FEDER funds under project MAT2011-28385.

%The financial support provided  is greatly appreciated

\section{Appendices}

\subsection{Appendix A: Alternative procedure for evaluation of Eqs. \ref{sigmak}, \ref{Pk}, \ref{promedioAk}, and \ref{funcionl}}

It is clear from Eq. \ref{promedioAk} that the integrals of the right hand side do not depend on $k$. Thus, the total cross-section, $\sigma_{k}$ can be written as \cite{Puertas1997}:
\begin{equation}\label{sigmak2}
    \sigma_{k}=k\sigma_{1}+\frac{4}{9}\pi a^{2} \alpha^{4} (m-1)^{2}\sum_{i,j;\; i \neq j}^{k} I_{ij},
\end{equation}
\begin{equation}\label{Iij}
    I_{ij}=I_{n}=\int_{0}^{\pi}\bigg \{\int...\int \rho(\gamma_{1},...,\gamma_{n}) \frac{sin hs_{ij}(\gamma_{1},...,\gamma_{n})}{hs_{ij}(\gamma_{1},...,\gamma_{n})}d\gamma_{1}...d\gamma_{n}\bigg\} \quad x \quad P_{1}(\vartheta)(1+cos^{2}\vartheta)sin\vartheta d\vartheta,
\end{equation}
where $\rho(\gamma_{1},...,\gamma_{n})$ is a flat distribution function, and $n$ is the number of particles between particles $i$ and $j$.

In the case that we consider a single particle ($n=-1$), Eq. \ref{Iij} can be written as \cite{Puertas1997}:
\begin{equation}\label{I-1}
    I_{-1}=\int_{0}^{\pi} P_{1}(\vartheta)(1+cos^{2}\vartheta)sin\vartheta d\vartheta.
\end{equation}

For the doublet, particles are directly connected and $n=0$, simplifying Eq. \ref{Iij} to:
\begin{equation}\label{I0}
    I_{0}=2\int_{0}^{\pi} \frac{\sin(hS_{12})}{hs_{12}}P_{1}(\vartheta)(1+cos^{2}\vartheta)sin\vartheta d\vartheta.
\end{equation}

For the case $n=1$, we have:
\begin{equation}\label{I1}
 I_{1}=2\int_{0}^{\pi} \int_{-2\pi/3}^{2\pi/3}\rho \frac{\sin(hs_{13}(\gamma_{1}))}{hs_{13}(\gamma_{1})}P_{1}(\vartheta)(1+cos^{2}\vartheta)sin\vartheta d\vartheta.
\end{equation}
Finally for $n=2$, we can write:
\begin{equation}\label{I2}
    I_{2}=2 \int_{0}^{\pi} \int_{-2\pi/3}^{2\pi/3} \int_{-2\pi/3}^{2\pi/3} \rho^{2} \frac{\sin(hs_{14}(\gamma_{1}, \gamma_{2}))}{hs_{14}(\gamma_{1}, \gamma_{2})}P_{1}(\vartheta)(1+cos^{2}\vartheta)sin\vartheta d\vartheta.
\end{equation}

Using the definition of $I_{n}$, the total cross-section of an aggregate of $N$ particles can be written as follows:
\begin{equation}\label{sigman}
    \sigma_{n}=N \sigma_{1}+\frac{4}{9}\pi a^{2} \alpha^{4} (m-1)^{2}\sum_{n=0}^{N-1} (N-n-1)I_{n},
\end{equation}
since there are $N - n - 1$ pairs of particles with $n$ particles between them. Here, $k_{CS}$ $I_{-1}= \sigma_{1}$ and $k_{CS}=(4/9) \pi a^{2} \alpha^{4} (m-1)^{2}$.

Now, using Smoluchowski kinetics \cite{Smoluchowski1917} and considering a unique kinetic constant for coagulation, $k_{f}$ independent of the aggregate form or size, the change in the number of aggregates with $k$ particles is given by:
\begin{equation}\label{Smoluchowski3}
    \frac{dn_{k}(t)}{dt}=\frac{1}{2}k_{f} \sum_{i,j=1, \; i+j=k}^{N} n_{i}(t)n_{j}(t)-k_{f}n_{k}(t) \sum_{i=1}^{\infty} n_{i}(t).
\end{equation}
Thus, $k$-particle aggregate concentration as a function of time, $n_{k}(t)$ can be obtained analytically \cite{Smoluchowski1917,Sonntag1987}
\begin{equation}\label{nk}
    n_{k}(t)=n_{0}\frac{t_{p}^{k-1}}{(1+t_{p})^{k+1}}    \qquad   t_{p}=k_{f}n_{0}t,
\end{equation}
where $n_{0}$ is the initial concentration of particles. The turbidity can be expressed as:
\begin{equation}\label{turbidity2}
    \tau=k_{CS} \; n_{0}\sum_{n=1}^{\infty} I_{n-2}\frac{t_{p}^{n-1}}{(1+t_{p})^{n-1}}.
\end{equation}

Puertas and de las Nieves \cite{Puertas1997} demonstrated that this expression retains the most important contributions to the summation if it is truncated at $n=2$.

\subsection{Appendix B: Balance of aggregates for a flocculating-coalescing emulsion}

In the case where only aggregation occurs, the balance of mass gives:
\begin{equation}\label{Definitionnk}
   \sum_{k} n_{k}=n_{total}.
\end{equation}

\begin{figure}[htbp]
\centering
   \includegraphics[width=5cm]{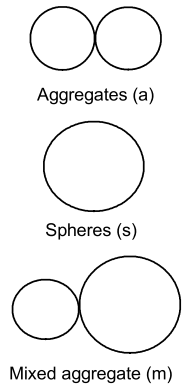}\\
  \caption{Types of aggregates formed in an emulsion.}\label{ASM}
\end{figure}

The particles can increase their size due to aggregation, coalescence, and aggregation of bigger drops with primary particles to form ``mixed'' aggregates Fig. \ref{ASM}.
\begin{equation}\label{definitionnt}
    n_{total}=n_{1}+\sum_{k=2} \left (n_{k,s}+n_{k,a}+n_{k,m}\right ),
\end{equation}
where, $n_{k,s}$ is the number of the drop resulting from coalescence, $n_{k,a}$ represents the number of aggregates with $k$ primary particles, and $n_{k,m}$ is the number of mixed aggregates resulting from the coalescence and aggregation of $k$ particles.

If both aggregates and spherical drops form as a result of binary collisions, they should be a fraction ($x$) of the total number of aggregates predicted by Smoluchowski for the case of irreversible flocculation:
\begin{equation}
  n_{k,s}= n_{k}x_{k,s},
\end{equation}
\begin{equation}
  n_{k,a} = n_{k}x_{k,a},
\end{equation}
\begin{equation}
  n_{k,m} = n_{k}x_{k,m}.
\end{equation}

Therefore,
\begin{equation}\label{nyx}
    n_{1}+\sum_{k=2} \left(n_{k,s}+n_{k,a}+n_{k,m}\right)= n_{1} + \sum_{k=2} n_{k}\left ( x_{k,s}+x_{k,a}+x_{k,m}\right).
\end{equation}
For each $k$:
\begin{equation}\label{sumadeX}
  x_{k,s}+x_{k,a}+x_{k,m}=1.
\end{equation}
Hence,
\begin{equation}\label{nyx2}
    n_{1}+\sum_{k=2} \left(n_{k,s}+n_{k,a}+n_{k,m}\right)= n_{1} + \sum_{k=2} n_{k} = n_{total}.
\end{equation}
That is, although the number of initial particles is different from the number of final particles, the total number of aggregates is the same for a flocculating suspension and a flocculating-coalescing emulsion. These two systems differ in the types of aggregates formed. When coalescence occurs, it generates mixed aggregates which are counted as aggregates of size $k$ in the absence of coalescence (their total volume corresponds to the sum of $k$ primary particles). Therefore, the number of aggregates does not change with respect to the definition of Smoluchowski, but their structure does.

Now, the turbidity can be redefined as:
\begin{equation}\label{Turbidity3}
    \tau = n_{1}\sigma_{1} + \sum_{k=2} x_{k,s} n_{k} \sigma_{k,s} + \sum_{k=2} x_{k,a} n_{k} \sigma_{k,a} + \sum_{k=2} x_{k,m} n_{k} \sigma_{k,m},
\end{equation}
\begin{equation}\label{Turbidity3-0}
    \tau = n_{1}\sigma_{1} + \sum_{k=2} n_{k} \left( x_{k,s} \sigma_{k,s} + x_{k,a} \sigma_{k,a} + x_{k,m}  \sigma_{k,m} \right).
\end{equation}

If $n_{k,m}$ is negligible ($x_{k,m}=0$), then
\begin{equation}\label{Turbidity3-1}
    \tau = n_{1}\sigma_{1} + \sum_{k=2} n_{k}\left(x_{k,s} \sigma_{k,s} + x_{k,a} \sigma_{k,a}\right).
\end{equation}

In this case, $x_{k,s}+x_{k,a}=1$, therefore Eq. \ref{Turbidity3-1} is equal to:
\begin{equation}\label{Turbidity3-2}
    \tau = n_{1}\sigma_{1} + \sum_{k=2} n_{k} \left[\left(1-x_{k,a}\right) \sigma_{k,s} + x_{k,a} \sigma_{k,a}\right].
\end{equation}

If all $x_{k,a}=x_{a}$ and $x_{k,s}=x_{s}$, then
\begin{equation}\label{Turbidity3-3}
    \tau = n_{1}\sigma_{1} + \sum_{k=2} n_{k} \left[\big(1-x_{a}\big) \sigma_{k,s} + x_{a} \sigma_{k,a} \right].
\end{equation}

If it is found that Eq. \ref{Turbidity3-3} is confirmed by the experiments but $x_{s}+x_{a}\neq 1$, this means that the average contribution of mixed aggregates is being incorporated into $\sigma_{k,a}$ and $\sigma_{k,s}$. If we equalize Eq. \ref{Turbidity3-2} to the exact form of the turbidity  (Eq. \ref{Turbidity3-0}), $\tau=\tau^{'}=\tau_{exp}$, then
\begin{equation}\label{Turbidity3-4}
    \sum_{k=2} n_{k} \left( x_{k,s} \sigma_{k,s} + x_{k,a} \sigma_{k,a} + x_{k,m}  \sigma_{k,m} \right)=\sum_{k=2} n_{k} \left( x_{k,s}^{'} \sigma_{k,s} +  x_{k,a}^{'} \sigma_{k,a}\right).
\end{equation}
Considering Eq. \ref{sumadeX}, we have:
\begin{equation} \label{Turbidity3-5}
     \sum_{k=2} n_{k} \left[ \sigma_{k,s} \left(x_{k,s}- x_{k,s}^{'}\right) + \sigma_{k,a} \left( x_{k,a}- x_{k,a}^{'}\right) \right] = - \sum_{k=2} n_{k} \sigma_{k,m} \left (1-x_{k,s}- x_{k,a} \right).
\end{equation}

It is clear then that the fact that $x_{s}+x_{a}\neq 1$ comes from mixed aggregates contributions. Moreover, this contribution can be evaluated since $x_{k,s}^{'}$ and $x_{k,a}^{'}$ are known. For this purpose, it is necessary to provide an approximate expression for $\sigma_{k,m}$.

\subsection{Appendix C: Calculation of errors}

In order to assign an error to the theoretical estimations of $k_{f}$ from models M1 to M4, an average error of the turbidity was approximated by the average difference between the experimental values of the turbidity ($\tau_{exp}$) and the theoretical prediction ($\tau_{fit}$).

The average error of the theoretical kinetic constant ($k_{f}$) was calculated from the deviates of $Abs$ ($\Delta Abs$)
\begin{equation}\label{Errortau}
    \Delta Abs= \frac{1}{N_{total}}\sum \left|Abs_{fit}-Abs_{exp} \right|,
\end{equation}
$N_{total}$ represents the number of values of $Abs$ for each experimental turbidity curve. The error of $k_{f}$ ($\Delta k_{f}$) was calculated from the propagation of errors of theoretical form of $\tau$ for each model, obtained using symbolic algebra (Mathematica 8.0.1.0):
\begin{equation}\label{Errork}
    \Delta k_{f}=  \frac{\Delta Abs}{\left | \partial \tau /\partial k_{f} \right |}
\end{equation}

Error bars of the initial slopes showed in Fig. \ref{PendienteSal} correspond to the standard deviation of three measurements. With these errors and the propagation of errors of Eq. \ref{Constante-dimeros}, $\Delta k_{11}$ was obtained.

%The other values of $W$ were calculating from the propagation of error of the \ref{W-Fuchs} and taking to account the errors of the kinetic constants show in \ref{kM1,kM3,kM4}.

The error of critical coagulation concentration ($\Delta CCC$) is given by:
\begin{equation}\label{ErrorCCC}
    \left | \frac{\Delta CCC}{CCC} \right |=\left | \frac{\Delta c}{c}\right | + \left |\frac{\Delta b}{b}\right |.
\end{equation}
The deviation of slope ($\Delta c$) and deviation of the intercept ($\Delta b$) were determined from the least-squares fitting of the $W$ vs. $[NaCl]$ curves.

\subsection{Appendix D: Comparison between Eq. \ref{sigmak} and Eq. \ref{sigmak2} (Model M1)}

\begin{figure}[htbp]
\centering
    \includegraphics[width=9cm]{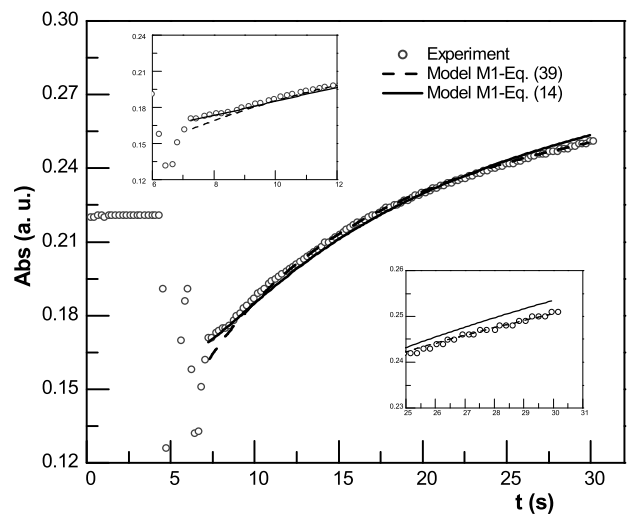}\\
  \caption{Evaluation of the model M1 using Eq. \ref{sigmak} and Eq. \ref{sigmak2}.}\label{ComparisonM1}
\end{figure}

A comparison between the two forms of evaluation of M1 is shown in Fig. \ref{ComparisonM1}. In the case of Eq. \ref{sigmak}, a value of $k_{max}=100$ was used. As can be seen, the adjustment of Eq. \ref{sigmak} to the experimental data is very good at the initial stage. However, the final points (long times) do not fit the data. On the contrary, Eq. \ref{sigmak2} fits very well in the final stage, while the initial points escape the fitting.

It is necessary to point out that the quality of the fitting depends on the values of the average radius of primary particles. It is well known that distinct average radii can be obtained using size distributions weighted in intensity, volume, area or number of particles. In the present calculations, we chose an ``optimal'' radius ($R_{fit}$) within the size range $R_{exp}\pm \Delta R_{exp} $ (Table \ref{Nanocharacteristics}) previously measured, in order to maximize the agreement between the experimental data and the theoretical models M3 and M4. Thus, in order to compare all the models on a similar basis, the same set of radii ($R_{fit}$) was considered to evaluate model M1 using Eqs. \ref{sigmak} and \ref{sigmak2}. Model M1 does not fit the DLCA data. The values of $k_{f}$ and $t_{0}$ for $380 \leq [NaCl] \leq 395 mM$ predicted from both equations are shown in Table \ref{Comparison}. It is clear that the results of Eq. \ref{sigmak} are of lower quality. Therefore, Eq. \ref{sigmak2} was used  for calculating the kinetics constants of model M1 (Table \ref{kM1}) under DLCA conditions.

\bibliographystyle{model3-num-names}

\newpage

\begin{table}[htbp]
 \centering
  \caption{Composition of the systems employed.}\label{Nanoemulsiones}
\begin{footnotesize}
 \vspace{0.3cm}
\begin{tabular}{c c c c}
  \hline
  Parameters &	Starting 	&  Concentrated & Final \\
   & system & nanoemulsion & nanoemulsion\\
  \hline
  $\phi$	& 0.84 &	$6.5 \times 10^{-2}$ &	$3.2 \times 10^{-4}$ \\

  $[SDS]$ (M)	& $0.35$ & 	$2.4 \times 10^{-2}$	& $8.0 \times 10^{-3}$\\

  $[NaCl]$ (M)	& 1.40	& $2.0 \times 10^{-2}$  &	$9.5 \times 10^{-5}$ \\

  $[Isopentanol]$ (wt\%)	&	6.50  &	$4.6 \times 10^{-1}$	& $ 2.2 \times 10^{-3}$\\

  Droplets/mL	& - &	$4.1 \times 10^{13}$ &	$2.0 \times 10^{11}$\\
  \hline
\end{tabular}
  \end{footnotesize}
\end{table}

\begin{table}[htbp]
 \centering
  \caption{Characterization of the nanoemulsions. }\label{Nanocharacteristics}
\begin{footnotesize}
 \vspace{0.3cm}
\begin{tabular}{c c c}
  \hline
  Mean radius (nm)	& 72.5 \\

  Standard deviation (nm)	& 25.4\\

  Coefficient of variation (CV)	& 35 \% \\

  $\zeta$ -potential (mV)	& -81.7 $\pm$ 2.5 \\
 \hline
\end{tabular}
  \end{footnotesize}
\end{table}

\begin{table}[htbp]
 \centering
  \caption{Values of $k_{11}$ from Eq. \ref{Constante-dimeros}.}\label{k11table}
\begin{footnotesize}
\vspace{0.3cm}
\begin{tabular}{c c c}
  \hline
  [NaCl]  &	$R_{exp}$ &  $ k_{11} $ 	 \\
 (mM) &	 (nm)&   ($m^{3} s^{-1}$)	 \\
  \hline
  380 	    &   75.2  &	  $(2.1 \pm 0.8) \times 10^{-19}  $   	  \\

  390 	    &   78.0  &   $(3.5 \pm 0.6) \times 10^{-19}  $      \\

  395	    &   74.3  &   $(3.8 \pm 1.9) \times 10^{-19}  $   	\\

  400 	    &	75.1  &   $(1.2 \pm 0.1)\times 10^{-18}  $      \\

  410	    &	78.0  &	  $(1.6 \pm 0.6) \times 10^{-18}  $      \\

  420	    &	76.1  &	  $(1.9 \pm 0.2) \times 10^{-18}  $     \\

  430	    &	74.3  &	  $(3.0 \pm 0.4) \times 10^{-18} $      \\

  450	    &	74.3  &	  $(2.8 \pm 0.3) \times 10^{-18}  $     \\

  475	    &	75.0  &	  $(3.2 \pm 0.6) \times 10^{-18}  $     \\

  500	    &	72.3  &	  $(3.3 \pm 0.3) \times 10^{-18}  $     \\

  550	    &	72.3  &	 $ (3.5 \pm 0.7) \times 10^{-18}  $     \\

  600	    &	75.0  &	  $(3.1 \pm 0.1) \times 10^{-18}  $     \\
  \hline
\end{tabular}
  \end{footnotesize}
\end{table}

\begin{table}[htbp]
 \centering
  \caption{Kinetic constants obtained from model M1.}\label{kM1}
\begin{footnotesize}
\vspace{0.3cm}
\begin{tabular}{c c c c c c}
  \hline
  [NaCl]  &	$R_{fit}$ &  $ k_{M_{1}}$  & $t_{0,exp}$  & $t_{0,fit}$ \\
  (mM) &	(nm)&  ($m^{3} s^{-1}$)&   (s)  & (s) \\
  \hline
  380 	 &   64.0  &   $(1.4 \pm 0.3)\times 10^{-19}$ & 11.9 & 11.0 \\

  390 	 &   64.0  &   $(3.3 \pm 0.2)\times 10^{-19}$ & 7.2 & 7.0 \\

  395	 &   65.0  &   $(3.3 \pm 2.5)\times 10^{-19}$ &  6.3 & 7.8 \\

  400   &	66.0  &	  $(4 \pm 18)\times 10^{-18}$ &   6.3 & 7.5 \\

  410  &	66.0  &	  $(4 \pm 38)\times 10^{-18}$ &	 6.2 & 6.5 \\

  420  &	75.0  &	  $(2 \pm 44)\times 10^{-17}$ &	 7.4 & 7.5 \\

  430  &	63.0  &	  $(6 \pm 70)\times 10^{-18}$ &	6.3  & 7.0\\

  450  &	69.3  &	  $(5 \pm 27)\times 10^{-18}$ &	6.5 &  6.7\\

  475  &	58.0 &	  $(5 \pm 35)\times 10^{-18}$ &	7.3 & 7.4 \\

  500  &	69.7  &	  $(7 \pm 2)\times 10^{-18}$ &	  6.4 & 6.6 \\

  550	 &	72.7  &	  $(7 \pm 29)\times 10^{-19}$ &	6.4 & 8.9 \\

  600	 &	57.7  &	  $(5 \pm 148)\times 10^{-19}$ &	7.3 & 7.8 \\
  \hline
\end{tabular}
  \end{footnotesize}
\end{table}

\begin{table}[htbp]
 \centering
  \caption{Kinetic constants obtained from model M3.}\label{kM3}
\begin{footnotesize}
\vspace{0.3cm}
\begin{tabular}{c c c c c c }
  \hline
  [NaCl] &	$R_{fit}$ &  $ k_{M3}$ & $g$ &   $t_{0,exp}$  & $t_{0,fit}$ \\
   (mM) &	(nm)&  ($m^{3} s^{-1}$)&  &   (s) &  (s)\\
  \hline
  380 	 &   64.0  &	  $(1.7 \pm 0.1)\times 10^{-19}$ &	 1.35 &	 11.9 & 13.0 \\

  390 	 &   64.0  &	  $(2.2 \pm 0.1)\times 10^{-19}$ &    1.07  &  7.2 & 6.6  \\

  395	 &   65.0  &	  $(3.5 \pm 0.3)\times 10^{-19}$ &    1.00 &	6.3 & 7.1   \\

  400   &	66.0  &	  $(6.7 \pm 0.2)\times 10^{-19}$ &    0.62 &	 6.3 & 7.2   \\

  410  &	66.0  &	  $(7.9 \pm 0.4)\times 10^{-19}$ &	  0.64  & 6.2 & 6.8 \\

  420  &	75.0  &	  $(2.2 \pm 0.1)\times 10^{-18}$ &	  0.69 &  7.4 & 7.6   \\

  430	 &	63.0  &	  $(1.0 \pm 0.1)\times 10^{-18}$ &	  0.78  &  6.3 & 7.1 \\

  450  &	69.3  &	  $(1.3 \pm 0.1)\times 10^{-18}$ &	  0.73 &	6.5 & 6.9   \\

  475  &	58.0  &	  $(8.4 \pm 0.2)\times 10^{-19}$ &	  0.79  &   7.3 & 7.5 \\

  500 &	69.7  &	  $(1.5 \pm 0.1)\times 10^{-18}$ &	   0.78  &  6.4 & 6.8\\

  550	 &	72.7  &	  $(1.5 \pm 0.1)\times 10^{-18}$ &	  0.81  &  6.4 & 8.9\\

  600	 &	57.7  &	  $(6.7 \pm 0.1)\times 10^{-19}$ &	   0.81  &  7.3 & 7.4\\
  \hline
\end{tabular}
  \end{footnotesize}
\end{table}

\begin{table}[htbp]
 \centering
  \caption{Kinetic constants obtained from model M4.}\label{kM4}
\begin{footnotesize}
\vspace{0.3cm}
\begin{tabular}{c c c c c c c}
  \hline
  [NaCl] &	$R_{fit}$ &  $k_{M4}$  & $p$ & $t_{0,exp}$  & $t_{0,fit}$ \\
   (mM) &	 (nm)&    ($m^{3} s^{-1}$)	&  &  (s) &  (s)\\
  \hline
  380 	 &   64.0  &      $(2.0 \pm 0.1)\times 10^{-19}  $  &	 1.68  & 11.9 & 13.1  \\

  390 	 &   64.0  &	  $(2.2 \pm 0.1)\times 10^{-19}  $  &	 1.11  & 7.2 & 6.6\\

  395	 &   65.0  &	  $(3.5 \pm 0.3)\times 10^{-19}  $  &	 1.00 & 6.3 & 7.1 \\

  400   &	66.0  &	     $(6.1 \pm 0.1)\times 10^{-19}  $  &	 0.43 & 6.3 & 7.2 \\

  410  &	66.0  &	     $(7.0 \pm 0.3)\times 10^{-19}  $  &	 0.45 & 6.2 & 6.8\\

  420  &	75.0  &	     $(1.9 \pm 0.1)\times 10^{-18}  $  &	 0.47 & 7.4 & 7.5\\

  430	 &	63.0  &	     $(9.0 \pm 0.2)\times 10^{-19}  $  &	 0.52 & 6.3 & 6.8 \\

  450  &	69.3  &	    $(1.2 \pm 0.1)\times 10^{-18}  $  &	 0.55 & 6.5 & 6.9 \\

  475  &	58.0  &	     $(7.3 \pm 0.1)\times 10^{-19}  $  &	 0.61 & 7.3 & 7.4\\

  500  &	69.7  &	    $(1.4 \pm 0.1)\times 10^{-18}  $  &	 0.60 & 6.4 & 6.9\\

  550	 &	72.7  &	    $(1.4 \pm 0.1)\times 10^{-18}  $  &	 0.67 & 6.4 & 8.9\\

  600	 &	57.7  &	    $(5.9 \pm 0.1)\times 10^{-19}  $  &	 0.65 & 7.3 & 7.3 \\
  \hline
\end{tabular}
  \end{footnotesize}
\end{table}

\begin{table}[htbp]
 \centering
  \caption{Values of critical coagulation concentration.}\label{CCC}
\begin{footnotesize}
 \vspace{0.3cm}
\begin{tabular}{c c c}
  \hline
  Data & CCC (mM NaCl)\\
   \hline
  From $W_{exp}$ &  412 $\pm$ 6\\
  From $W_{k_{11}}$ &  411 $\pm$ 3\\
  From $W_{M3}$ 	&  415 $\pm$ 9\\
  From $W_{M4}$ 	&  418 $\pm$ 13\\
  \hline
\end{tabular}
  \end{footnotesize}
\end{table}

\begin{table}[htbp]
 \centering
  \caption{Comparison between the values of $k_{f}$ and $t_{0}$ for $380 \leq [NaCl] \leq 395$ mM using Eq. \ref{sigmak2} and Eq. \ref{sigmak}.}\label{Comparison}
\begin{footnotesize}
 \vspace{0.3cm}
\begin{tabular}{l c c c c c}
  \hline
  $[NaCl]$  & $k_{f}$-Eq. \ref{sigmak}  & $t_{0,fit}$ & $k_{f}$-Eq. \ref{sigmak2}  & $t_{0,fit}$ & $t_{0,exp}$ \\
   (mM) &  ($m^{3} s^{-1}$) & (s) & ($m^{3} s^{-1}$)  &  (s)& (s) \\
   \hline
 380   &  $7.52 \times 10^{-20}$ & 8.3  & $1.37 \times 10^{-19}$ & 10.9  &  11.9\\
 390   &  $1.93 \times 10^{-19}$ & 6.4   & $3.32 \times 10^{-19}$ & 7.0   &  7.2\\
 395   &  $3.82 \times 10^{-19}$ & 6.6   & $3.30 \times 10^{-19}$ & 7.8   &  6.3\\
  \hline
\end{tabular}
  \end{footnotesize}
\end{table}

\newpage

\textbf{Figure Captions}

\begin{description}
 \item [Figure 1:] Definition of $\gamma$ for a chain of three particles (a) and four particles (b).
 \item [Figure 2:] [NaCl] vs. [SDS] phase diagram of the \emph{water/NaCl/isopentanol/SDS/dodecane} system at $6.5$ \%wt isopentanol and $f_{w}=0.20$.
 \item [Figure 3:] Droplet size distributions of three independent nanoemulsions synthesized from three different mother emulsions measured with a Malvern Zetasizer 2000.
 \item [Figure 4:] Evolution of $R_{exp}$ as a function of time for several nanoemulsions.
 \item [Figure 5:] Typical curve of $Abs$ vs. $t$.
 \item [Figure 6:] Aggregation kinetics of the nanoemulsion.
 \item [Figure 7:] Results of two measurements of the initial droplet size distribution of the same nanoemulsion using a BI-200SM  Goniometer: (a) Contin, (b) Cumulants analysis.
 \item [Figure 8:] Temporal evolution of the droplet size distributions for a nanoemulsion with 380 mM NaCl using a BI-200SM Goniometer: (a) Contin, (b) Cumulants analysis.
 \item [Figure 9:] Temporal evolution of the droplet size distributions for a nanoemulsion with 600 mM NaCl using a BI-200SM Goniometer: (a) Contin, (b) Cumulants analysis.
 \item [Figure 10:] Droplet size distributions after 1 min of addition of different concentrations of salt using a BI-200SM Goniometer: (a) Contin, (b) Cumulants analysis.
 \item [Figure 11:] Comparison of the theoretical curves of absorbance predicted by models M1 and M2 with the experimental measurements of slow aggregation ($[NaCl]=390$ mM).
 \item [Figure 12:] Comparison of the theoretical curves of absorbance predicted by models M3 and M4 with the experimental measurements of slow aggregation ($[NaCl]=390$ mM).
 \item [Figure 13:] Comparison of theory and experiment for the case of fast aggregation ($[NaCl]=600$ mM).
 \item [Figure 14:] Predictions of model M1 for $k_{max}= 2, 3, 4$ and $100$ ($[NaCl]=390$ mM).
 \item [Figure 15:] Comparison between $W_{k_{11}}$ and $W_{exp}$.
 \item [Figure 16:] Comparison between theoretical and experimental values of $W$.
 \item [Figure 17:] Comparison of the behavior of $\tau_{aprox}$ with $\tau (R_{fit})$ for models M1 and M3 at $[NaCl]=390$ mM. (The number of data points of the experimental curve was reduced in order to highlight the quality of the fittings).
 \item [Figure 18:] Comparison of the behavior of $\tau_{aprox}$ with $\tau (R_{fit})$ for models M1 and M3 at $[NaCl]=475$ mM. (The number of data points of the experimental curve was reduced in order to highlight the quality of the fittings).
 \item [Figure 19:] Types of aggregates formed in an emulsion.
 \item [Figure 20:] Evaluation of the model M1 using Eq. \ref{sigmak} and Eq. \ref{sigmak2}.
\end{description}

\newpage

\clearpage

\end{document}